\newcommand{\be}{\begin{equation}}
\newcommand{\bea}{\begin{eqnarray}}
\newcommand{\ee}{\end{equation}}
\newcommand{\eea}{\end{eqnarray}}
\newcommand{\qa}{\alpha}
\newcommand{\qb}{\beta}
\newcommand{\qg}{\gamma}
\newcommand{\qG}{\Gamma}
\newcommand{\qd}{\delta}
\newcommand{\qD}{\Delta}
\newcommand{\qe}{\varepsilon}

\newcommand{\qh}{\eta}

\newcommand{\qk}{\kappa}

\newcommand{\qL}{\Lambda}

\newcommand{\qr}{\rho}
\newcommand{\qs}{\sigma}

\newcommand{\qt}{\tau}
\newcommand{\qf}{\varphi}

\newcommand{\qo}{\omega}
\newcommand{\qO}{\Omega}

\newcommand{\tr}{{\rm tr}\;}
\newcommand{\Tr}{{\rm Tr}\;}

\newcommand{\inv}{^{-1}}
\newcommand{\st}{^{*}}
\newcommand{\dagg}{^{\dag}}
\newcommand{\prt}{\partial}

\newcommand{\intd}[1]{\int \!\! d{#1} \;}

\newcommand{\intdxx}{\int \!\! d^{2}x \;}

\newcommand{\pathint}[1]{\int\!\! {\cal D} #1 \;}

\newcommand{\fr}[2]{{\textstyle \frac{#1}{#2}}}
\newcommand{\half}{\mbox{$\textstyle \frac{1}{2}$}}

\newcommand{\kwart}{\mbox{$\textstyle \frac{1}{4}$}}
\newcommand{\naar}{\rightarrow}

\newcommand{\expec}[1]
{\raisebox{.4ex}{\scriptsize $\langle$} #1
 \raisebox{.4ex}{\scriptsize $\rangle$}}

\newcommand{\nn}{\nonumber}

\newcommand{\scez}{\setcounter{equation}{0}}
\newcommand{\ns}{\scez\section}
\renewcommand{\theequation}{\thesection .\arabic{equation}}

\newcommand{\eff}{_{\rm eff}}

\newcommand{\tI}{\tilde{I}}
\newcommand{\Ian}{{\rm I}^\alpha_n}
\newcommand{\Iamn}{{\rm I}^\alpha_{-n}}

\documentstyle[preprint,aps,eqsecnum,epsf,prb,tighten]{revtex}
\begin{document}

\draft

\title{(Mis-)handling gauge invariance \\ in the theory of the quantum Hall
 effect  II: \\ Perturbative results}

\author{M.A. Baranov\cite{Misha}, A.M.M. Pruisken, B. \v{S}kori\'{c}}

\address
{Institute for Theoretical Physics, University of Amsterdam,
Valckenierstraat 65, \\ 1018 XE Amsterdam, The Netherlands}

\date{October 9, 1998}
\maketitle

\begin{abstract}
\noindent
The concept of ${\cal F}$-invariance, which previously arose
in our analysis of the integral and half-integral quantum Hall
effects, is studied in $2\! +\! 2\qe$ spatial dimensions.\newline
We report the results of a detailed renormalization group analysis and
establish the renormalizability of the (Finkelstein) action to two
loop order. We show that the infrared behaviour of the theory can be
extracted from gauge invariant (${\cal F}$-invariant) quantities
only. For these quantities (conductivity, specific heat) we derive
explicit scaling functions.\newline
We identify a bosonic quasiparticle density of states which develops a
Coulomb gap as one approaches the metal-insulator transition from the
metallic side. We discuss the consequences of ${\cal F}$-invariance
for the strong coupling, insulating regime.
\end{abstract}


\ns{Introduction}
\subsection{General introduction}

Two of the authors recently proposed a topological extension of the
Finkelstein non-linear sigma model for localization and interaction
effects \cite{PruiskenBaranov}. This theory was motivated by the extended experimental work
on scaling in the quantum Hall regime by H.P. Wei {\it et al.}
\cite{a8,a10,a11}. It
was shown that the interacting electron gas shares many of the features which
were previously found for free electrons in a magnetic field. In
particular, the same scaling diagram was obtained, indicating that the
topological concept of {\em instanton vacuum}, rather than being a
free particle theory of the plateau transitions alone, presumably has
a much more profound significance in the theory of quantum transport.

In a previous paper \cite{3A} we elaborated on the microscopic origins of the
effective sigma model action. We recognized a new symmetry in the
problem (`${\cal F}$-invariance') which previously has gone largely unnoticed and
which is intimately related to the electrodynamic $U(1)$ gauge
invariance of the theory. 

For ordinary metallic conductors the consequences of ${\cal F}$-invariance may be
summarized by saying that the Einstein relation between conduction and
diffusion no longer describes the process of quantum
transport. Instead, the internally generated electric field due to the
Coulomb interactions is what enters into the transport equations. We have
shown that this aspect of ordinary metals has direct consequences for
the composite fermion approach to the half-integral quantum Hall
effect.

The half-integral effect, along with the transport problem in weak
magnetic fields, is described by our topological action by working
with the Chern-Simons or external fields in {\em weak coupling} or
tree level approximation. On the other hand, in subsequent papers
\cite {b31} we
will address the Luttinger liquid theory of the chiral fractional
quantum Hall edge states \cite{b27}. The Luttinger liquid can be microscopically
obtained from our topological action by working in the opposite limit
of {\em strong coupling}.

The Fermi-liquid state of the half-integral effect and the Luttinger
liquid state of the edges are completely different physical scenarios
which are actively and separately being pursued in the recent
literature on the quantum Hall effect. Since they appear as the
extreme {\em weak} and {\em strong coupling} limits of a single
topological action, it becomes possible to pursue a more ambitious
program and see how they are related.

\vskip0.5cm

In this paper we report the results of detailed renormalization group
studies of the perturbative weak coupling regime. Our main objective
is to further investigate the fundamental consequences of ${\cal F}$-invariance
and to establish the renormalizability of the theory to two loop
order. 
In section~\ref{secbackground} we present a different version of
Finkelstein's momentum shell computations \cite{b8}, 
namely the background field
method in dimensional regularization, which leads to major
computational advantages. One of the most important consequences of
the renormalization group procedure, however, is that the infrared
behaviour of the theory can be extracted from ${\cal F}$-invariant
quantities or correlation functions only. This aspect of the problem
has till now remained completely unnoticed.


For arbitrary correlation functions or renormalization group
procedures which are not ${\cal F}$-invariant, the perturbative expansions are
generally plagued by infrared problems which can not be resolved by
the renormalization group. This means, for example, that the expansion
procedure can not be used to show that the diffusion propagator (which
is {\em not} an ${\cal F}$-invariant quantity) contains an infrared cutoff of
the type $1/\qt_{\rm in}$, i.e. the inelastic scattering length as
naively obtained from the Golden Rule. For the same reason it is also
fundamentally incorrect to interpret the theory in terms of a
`Fermi-liquid with length scale dependent parameters' \cite{b35}.

\vskip0.5cm

Examples of ${\cal F}$-invariant quantities are the linear response formulae
which involve current and density correlation functions and, of course,
the free energy or grand canonical potential itself. The most
important results of this work are reported in Sections \ref{seclinresp} and
\ref{secfree_energy} where we compute these quantities and in
Section~\ref{secscaling} where we derive explicit scaling functions
for them.

Based on a two-loop expansion for the free energy
(Section~\ref{sectwoloop} and the Appendix) we are able to identify a
new quantity in the problem, namely a bosonic quasiparticle density of
states which enters the expression for the specific heat
(Section~\ref{secspecheat}). This
quantity develops a Coulomb gap as one approaches the metal-insulator
transition in $2\! +\! 2\qe$ dimensions from the metallic side
(Section~\ref{secCoulquasi}). 

\vskip0.5cm

As a logical follow-up we next discuss the problem of the strong
coupling, insulating phase (Section~\ref{secstrongcoupl}). 
We show that important
progress can be made by relying on the more familiar but completely
analogous theory of the classical Heisenberg ferromagnet. Our results
indicate that the insulating phase is dominated by additional terms in
the action which are usually dismissed as being `irrelevant'. This,
then, will serve as a starting point for an extended renormalization
group program. Progress along these lines, along with the consequences
of the Coulomb interactions for the plateau transitions in the quantum
Hall regime, will be reported elsewhere.

\subsection{Preliminaries}
\label{secprelim}
\subsubsection{Formalism}
\label{secformalism}

\begin{figure}
\begin{center}
\setlength{\unitlength}{1mm}
\begin{picture}(128,40)(0,0)
\put(0,5)
{\epsfxsize=3.5cm{\epsffile{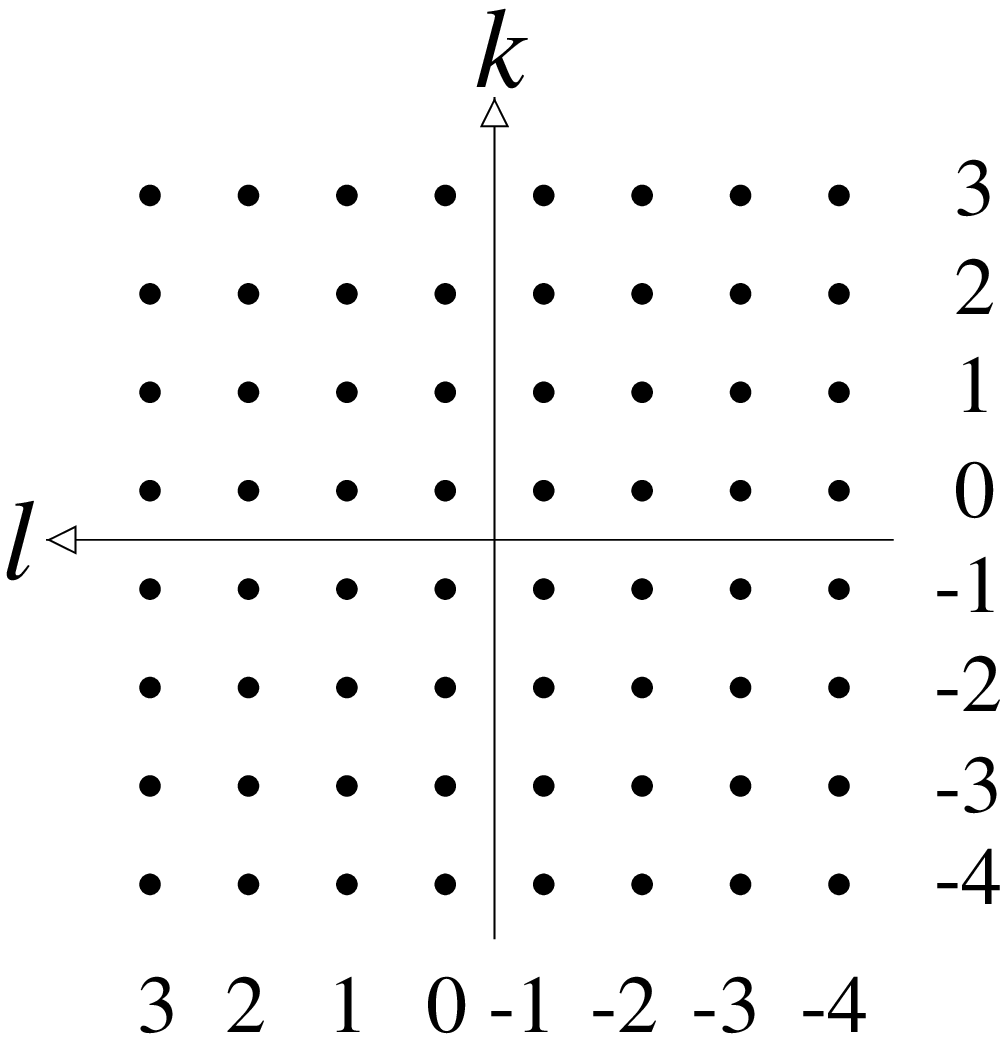}}}
\put(48,5)
{\epsfxsize=3.5cm{\epsffile{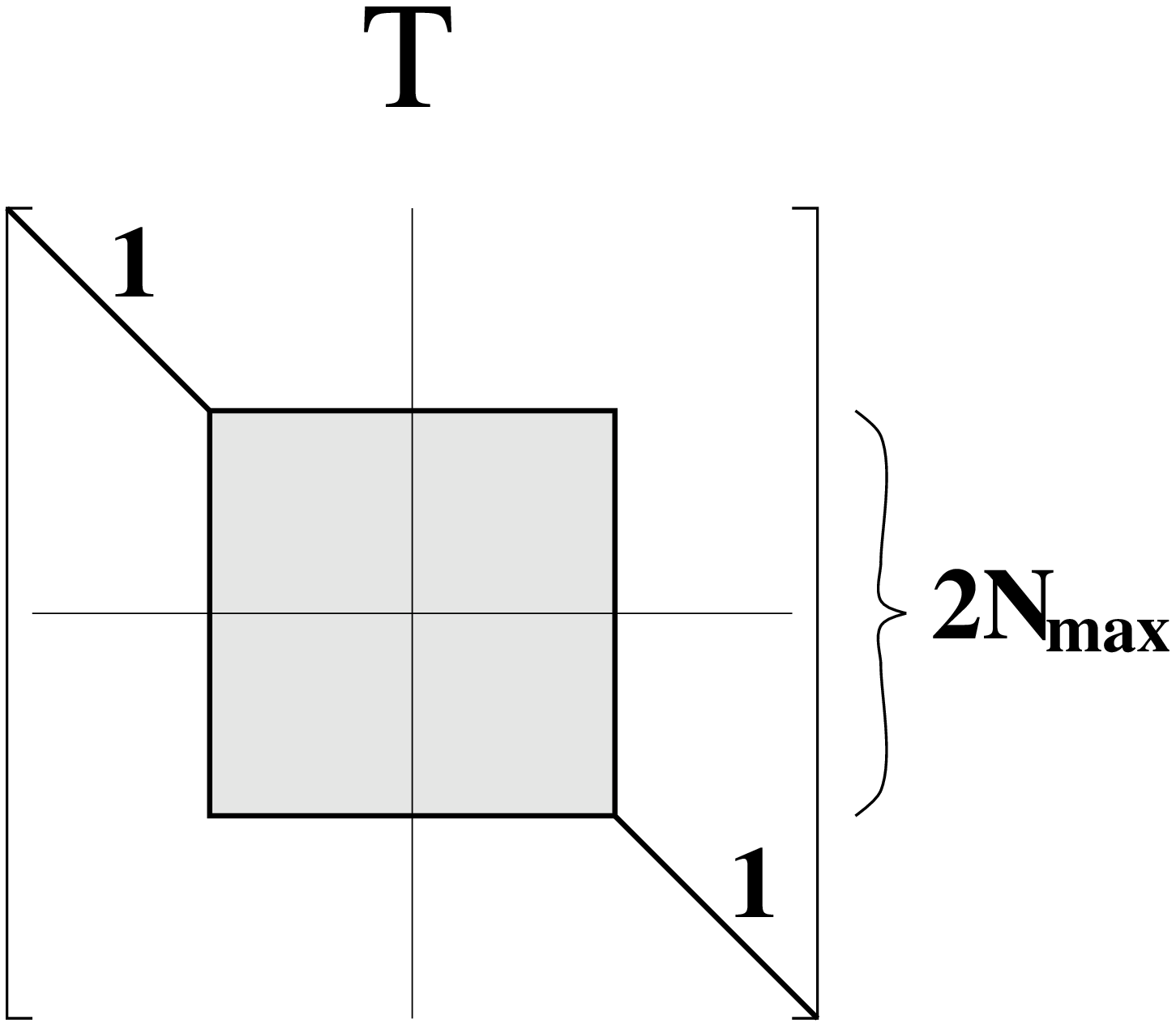}}}
\put(93,5)
{\epsfxsize=3.5cm{\epsffile{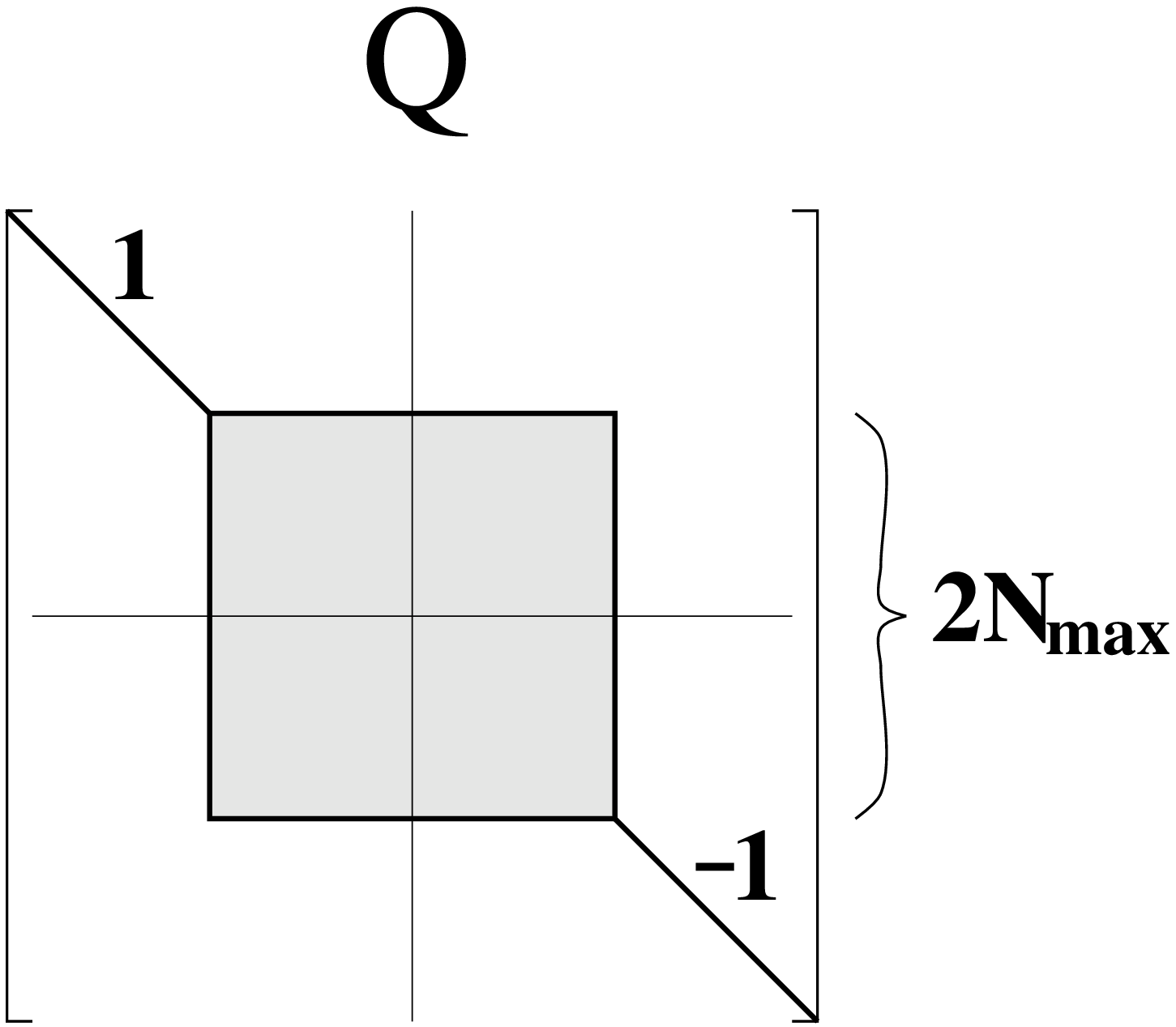}}}
\end{picture}
\caption{Sketch of a matrix $[\cdots]_{kl}$ with $k,l$ denoting the
Matsubara frequency indices. The `small' matrices $T$ and $Q$ are
nonzero only on the diagonal and within the square of size $2N_{\rm
max}\!\times\!2N_{\rm max}$ (shaded area).}
\label{figmatrix1TQ}
\end{center}
\end{figure}

In this and the following section we summarize the main results of the
$Q$-field theory for the quantum Hall effect as discussed in great
detail in our previous work \cite{3A}.
The electronic degrees of freedom in this theory
are contained in a matrix field $Q$, for which the following effective
action was derived,
\bea
	Z[A]=\pathint{Q}e^{S[Q,A]} \hskip0.5cm &;& \hskip0.5cm
	S[Q,A]=S_\qs[Q,A]+S_{\rm F}[Q]+S_{\rm U}[Q,A]
\label{SQA}
\eea
\bea
\label{Ssigma}
        S_\qs[Q,A] &=& -\fr{1}{8}\qs^0_{xx}\Tr[\prt_j-i\hat A_j, Q]
	[\prt_j-i\hat A_j, Q]
        \nn \\ &&
	-\fr{1}{8}\qs^0_{xy}\Tr \qe_{ij}Q
	[\prt_i-i\hat{A}_i,Q][\prt_j-i\hat A_j,Q] \\
\label{SF}
        S_{\rm F}[Q] &=& z_0\fr{\pi}{\qb} \left[
        {\sum_{n\qa}}\!\intdxx(\tr \Ian Q) (\tr \Iamn Q)
	+4\Tr\qh Q -6\Tr \qh\qL\right] \\
        S_{\rm U}[Q,A] &=& -\fr{\pi}{\qb}{\sum_{n\qa}}\intd{^2 x d^2 x' }
        [\tr \Ian Q-\fr{\qb}{\pi}(A_\qt)^\qa_{-n}
	+\fr{i\qb}{\pi\qr}\qs_{xy}^{II}B^\qa_{-n}](\vec x)
	U\inv(\vec x,\vec x') \times \nn\\ && \times
        [\tr \Iamn  Q-\fr{\qb}{\pi}(A_\qt)^\qa_n
	+\fr{i\qb}{\pi\qr}\qs_{xy}^{II}B^\qa_n](\vec x')
	-\fr{\qb}{4\pi\qr}(\qs_{xy}^{II})^2\intdxx B\dagg B.
\label{SU}
\eea
The various symbols appearing in this action have the following
meaning:
The $Q$ and $\Ian$ are matrices carrying combined replica (upper,
Greek) and Matsubara
frequency (lower, Latin) indices. For instance, the expression 
$\tr \Ian Q$ stands for 
$\sum_{kl}\sum_{\qb\qg} (\Ian)^{\qb\qg}_{kl} Q^{\qg\qb}_{lk}$.
The size of all matrices is 
$2N_{\rm max}'\!\!\times\!2N_{\rm max}'$ in frequency space; the frequency
indices run from $-N_{\rm max}'$ to $N_{\rm max}' \! -\! 1$, while the
replica indices run from $1$ to $N_r$. (See figure~\ref{figmatrix1TQ} for
the way in which we sketch matrices in frequency space.)
The $Q$-matrix is of the following form,
\bea
	Q=T\inv \qL T \hskip0.5cm &;& \hskip0.5cm	
	T\in SU(2N)
\label{defQ}
\eea
where $N\! =\! N_r\cdot N_{\rm max}$ with $N_{\rm max}\ll N_{\rm max}'$
(see figure~\ref{figmatrix1TQ}),
and $\qL$ is given by 
\be
	\qL^{\qa\qb}_{kl}=\qd^{\qa\qb}\left[
	\matrix{1 & 0\cr 0 & -1}\right]_{kl}.
\ee
The $\Ian$ lives in the $\qa$'th replica channel, while in
frequency space it is the unity matrix shifted by $n$ places
\be
	(\Ian)^{\qb\qg}_{kl}=\qd^{\qb\qa}\qd^{\qg\qa}\qd_{k-l,n}.
\ee
The `hat' ($\widehat{\hphantom{w}}$) 
on the external field $A$ in (\ref{Ssigma})
denotes a summation with the I-matrices,
\be
	\hat x :=\sum_{\qa n} x^\qa_n \Ian.
\ee
The external field is chosen such that $(A_\mu)^\qa_n$ is nonzero in
the same $n$-interval as $\tr \Ian Q$.
The matrix $\qh$ is given by
\be
	\qh^{\qa\qb}_{nm}=n \qd^{\qa\qb}\qd_{nm}.
\ee
The $\Tr$ stands for a matrix trace combined with spatial integration.
The $\qs_{ij}^0$ are the mean field conductances.
The $z_0$ is the singlet interaction amplitude
and $U\inv(\vec x,\vec x')$ is related to the screened Coulomb
interaction.
In momentum space $U\inv$ is given by
\be
	U\inv(q)=\fr{\pi/2}{\qr\inv+U_0(q)}
\ee
with $\qr\!=\! dn/d\mu$ the thermodynamic density of states.

\begin{figure}
\begin{center}
\setlength{\unitlength}{1mm}
\begin{picture}(90,45)(0,0)
\put(0,5)
{\epsfxsize=3.5cm{\epsffile{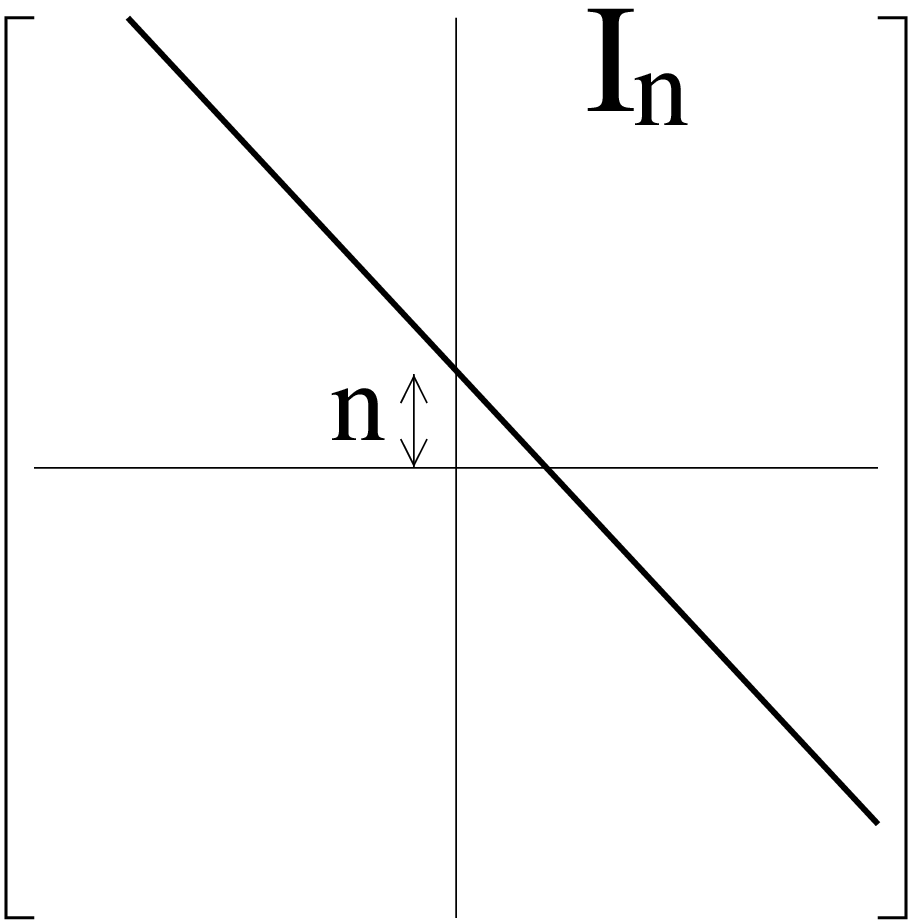}}}
\put(55,5)
{\epsfxsize=3.5cm{\epsffile{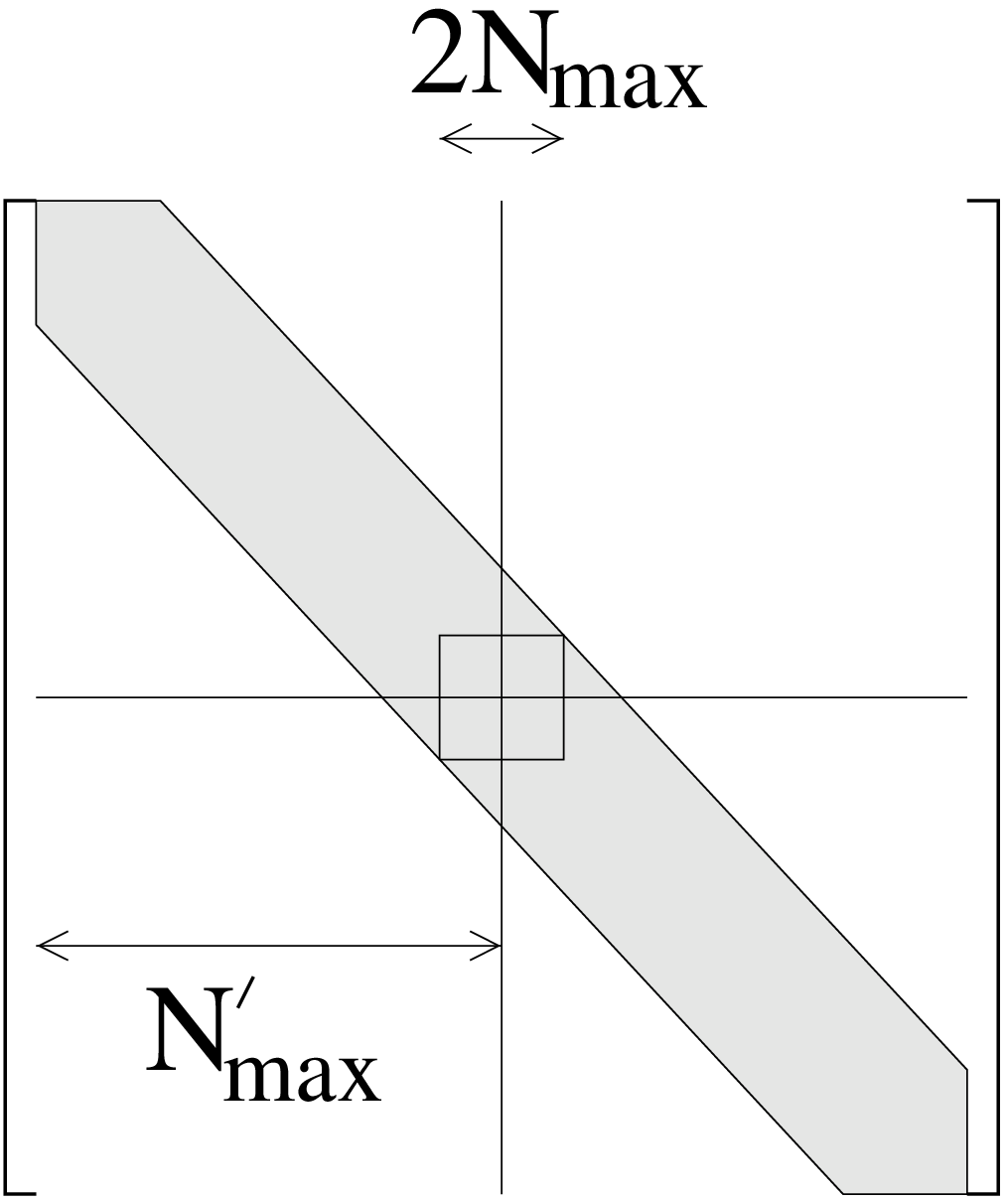}}}
\end{picture}
\caption{Left: structure of the `large' matrix $\Ian$ ($n\!\! > \!\! 0$).
Right: the summation interval
$n\! \in \! \{-2N_{\rm max}\! +\! 1,\cdots,2N_{\rm max}\! -\! 1 \}$
is indicated by the shaded area.}
\label{figIrange}
\end{center}
\end{figure}

\subsubsection{Gauge invariance}
\label{secgaugeinv}
Several important remarks have to be made with respect to the gauge
invariance of the theory (\ref{SQA}). 
At the first stages of the derivation of this action, all matrices
were infinite in size 
($N_{\rm max}\! =\! N_{\rm max}'\! =\!\infty$). The
infinite $\tI$-matrices obeyed 
$\tI^\qa_n \tI^\qb_m \! =\!\qd^{\qa\qb}\tI^\qa_{n+m}$,
forming an abelian algebra and generating the electromagnetic gauge
transformations,
\bea
	A_\mu\naar A_\mu+\prt_\mu\chi \hskip0.5cm &;& \hskip0.5cm
	Q\naar \exp\left[i\!\!\!\sum_{\qa,n=-\infty}^\infty 
	\!\!\!\!\chi^\qa_n \tI^\qa_n\right]\;
	Q\;\exp\left[-i\!\!\!\sum_{\qa,n=-\infty}^\infty 
	\!\!\!\!\chi^\qa_n \tI^\qa_n\right].
\label{infgauge}
\eea
At a certain point the cutoff $N_{\rm max}'$ had to be introduced,
destroying the $U(1)$ nature of the $I$-matrices. The truncated $\Ian$ do
not even span an algebra any more. The commutators are given by
\bea
        (\Ian {\rm I}^\qb_m)^{\mu\nu}_{kl}=(\tI^\qa_n \tI^\qb_m)^{\mu\nu}_{kl}
        g_{l+m} \hskip0.5cm &;& \hskip0.5cm
        [\Ian, {\rm I}^\qb_m]^{\mu\nu}_{kl}=\qd^{\qa\qb\mu\nu}\qd_{k-l,m+n}
        (g_{l+m}\!-g_{l+n})
\label{I,I}
\eea 
where $\qd^{\qa\qb\mu\nu}$ means that all replica indices have to be the
same, and $g_i$ is a step function equal to one if 
$i\in\{ -N_{\rm max}',\ldots,N_{\rm max}'\!\! -\! 1   \}$ 
and zero otherwise.
At this stage, there seems to be little hope of preserving any kind of gauge
invariance in the theory.

Disaster is averted in a very subtle way, however. By introducing the
second cutoff $N_{\rm max}\!\ll N_{\rm max}'$ for the matrix $T$ (and
thereby $Q$), most of the nasty aspects of the commutations (\ref{I,I}),
living at the edges of Matsubara frequency space, can be avoided.
The action (\ref{SQA}) is invariant under the truncated equivalent of
(\ref{infgauge}), 
\bea
	A_\mu\naar A_\mu+\prt_\mu\chi \hskip0.5cm &;& \hskip0.5cm
	Q\naar e^{i\hat\chi} Q e^{-i\hat\chi}.
\label{truncgauge}
\eea
The invariance can be checked using the following transformation rules
\bea
\label{transfI}
	\tr[\Ian \;e^{i\hat\chi}Qe^{-i\hat\chi}] &=&
	\tr \Ian Q+\fr{\qb}{\pi}(\prt_\qt\chi)^\qa_{-n} \\
\label{transfeta}
	\tr[\qh \;e^{i\hat\chi}Qe^{-i\hat\chi}] &=&
	\tr\qh Q-\fr{\qb}{2\pi}\tr Q\widehat{\prt_\qt\chi}
	-(\fr{\qb}{2\pi})^2 
	\sum_{n\qa}(\prt_\qt\chi)^\qa_{-n}(\prt_\qt\chi)^\qa_n.
\eea
The remarkable aspect of equations (\ref{transfI}, \ref{transfeta}) is
that they are {\em exact to all powers in $\chi$} as long as 
$N_{\rm max}\!\ll N_{\rm max}'$, and that the cutoff $N_{\rm max}'$ does
not appear in them, allowing one to send it safely to infinity.

Using (\ref{transfI}, \ref{transfeta}) it is easily seen that $S_{\rm
F}$ (\ref{SF}), the `Finkelstein' part of the action, is by itself invariant
under (\ref{truncgauge}). This fact is going to be very important in
section~\ref{secbackground}.
It can also be checked that (\ref{SU}) and the two terms in
(\ref{Ssigma}) are separately invariant. The invariance of the
$\qs_{xy}$-term holds as long as the sample has no
boundaries. The effect of a boundary on the theory is very interesting
and will be discussed in a subsequent paper.

We will often call the manipulations with the truncated I-matrices by
the name of `${\cal F}$-algebra' and denote the invariance
by `${\cal F}$-invariance'.
The invariance of the action does not automatically guarantee
invariance of the whole theory; since $Q$ and 
$e^{i\hat\chi}Qe^{-i\hat\chi}$ do not belong to the same manifold, we
can not absorb the $e^{i\hat\chi}$-rotation into the measure of the
$Q$-integration. The idea behind our approach is, however, that full
invariance is regained after sending the cutoff $N_{\rm max}$ to infinity.
In the sections that follow, it is always understood that this limit
is taken at the end of all calculations.

\ns{Background field renormalization}
\label{secbackground}
\subsection{The background field method}
\label{secbackmethod}

Let us start with the action (\ref{SQA}) without the external fields
$A_\mu$. We drop the topological term, since it is not going to
contribute to perturbation theory.
At the moment we are not interested in the full significance of the $S_{\rm
U}$ term, but only in the low-momentum limit.
Since $S_{\rm U}$ really stands for a higher dimensional operator, it
is preferable to work with a simpler theory in which the $U\inv$ in
(\ref{SU}) is
replaced by a $\qd$-function. This gives, up to a constant, 
\be
\label{Sback}
	S[Q] = -\fr{\qs_0}{8}{\rm Tr}(\nabla Q)^2
	+\fr{\pi}{\qb} z_0\int_x\left\{ c_0\sum_{\alpha n}{\rm tr}
	(\Ian Q){\rm
	tr}(\Iamn Q)+4{\rm tr}(\eta Q)\right\}.
\ee
We have introduced a parameter $c_0$ such that the
theory interpolates between the Coulomb case ($c_0=1$) and the free particle
case ($c_0\! =\! 0$).
Let us now define
\be
	\widehat Q=T_0^{-1}QT_0, 
\label{hatQ}
\ee
where $T_0$ is a `small' but
fixed and slowly varying background field of size, say, 
$2n_{\rm \max}\!\times 2n_{\rm \max}$, where 
$n_{\rm \max}\!\ll\! N_{\rm \max}$. 

If one pursues momentum shell computations then (\ref{hatQ}) is
loosely interpreted as a change of variables, where $T_0$ stands
for the `slow' modes which
should be kept. The $Q$ in (\ref{hatQ}) then really represents the
`fast' modes which should be eliminated. 
This `change of variables' idea is clearly
somewhat cavalier and complications arise in pursuing the theory
beyond one-loop order.

It is the purpose of this section however to show that the basic idea can be
put to work and we shall extract important information from it. More
specifically, if we keep working with a fixed but
`small' background field, then we can employ the more powerful method of
dimensional regularization and compute the effective action
for the $T_0$-field insertion. 
\be
	e^{S\eff[T_0]}=\pathint{Q}e^{S[Q,T_0]}.
\ee
As was shown in detail in the
context of the ordinary $\sigma$ model \cite{PruiskenWang}, 
this results in a very
effective way of extracting the pole terms in $\varepsilon$, i.e. the renormalization group coefficients ($Z$) of
the theory in $2\! +\! 2\qe$ dimensions.

\begin{figure}
\begin{center}
\setlength{\unitlength}{1mm}
\begin{picture}(100,65)(0,0)
\put(0,0)
{\epsfxsize=100mm{\epsffile{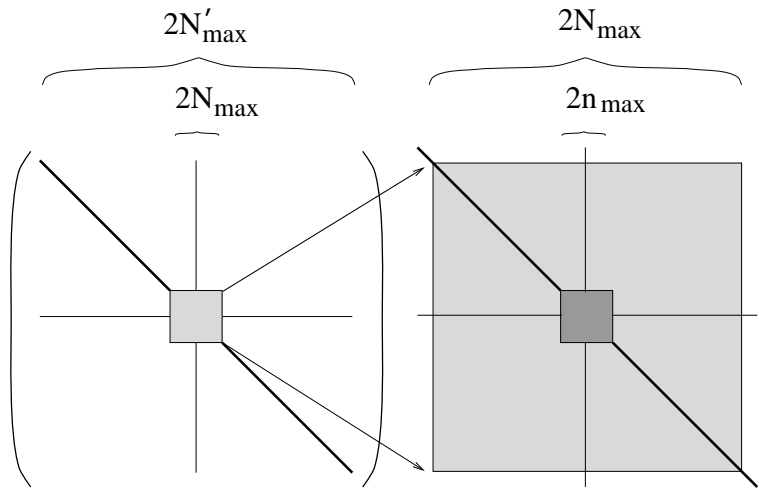}}}
\end{picture}
\caption{The relative sizes of $Q$ and the background field matrix
$Q_0$; see text.}
\label{figrg1}
\end{center}
\end{figure}

In addition to this we can give precise meaning to the idea of 'small'
background field $T_0$. In particular, since the $Q$ variables are
invariant under a local $U(N)\!\times\! U(N)$ transformation 
($N\! =\! N_{{\rm \max }}\cdot N_r$) we may conclude that the abovementioned
$S\eff[T_0]$
can be expressed in the local quantity
\be
\label{Q0}
	Q_0=T_0^{-1}\Lambda T_0. 
\ee
This quantity is of the same form as the original variable $Q$
(\ref{defQ}) except that the sizes $N_{\rm \max}'$, $N_{{\rm \max }}$ are
now replaced by $N_{{\rm \max }}$ and $n_{{\rm \max }}$ respectively. 
(See figure~\ref{figrg1}.)

In
different words, the procedure provides the additional information on the
renormalization of operators which follows from the requirement that the
form of the original action and the form of the shift in the free
energy (which we call `effective action' in $Q_0$) are the same.
Although the basic idea is quite straightforward, it is important to stress
that in practice the scheme turns out to be extremely rich and subtle. In
order to appreciate the flow of information that can be extracted from it we
proceed and describe the results in a step by step fashion. 
For reasons to be explained shortly, we add a
$U(N)\!\times\! U(N)$ invariant regulator proportional to $h_0^2$. The
action becomes
\bea
\label{Sbackground}
	S[Q] &=& -\fr{\qs_0}{8}{\rm Tr}(\nabla Q)^2
	+\fr{\pi}{\qb} z_0\int_x\left\{ c_0\sum_{\alpha n}{\rm tr}
	(\Ian Q){\rm
	tr}(\Iamn Q)+4{\rm tr}(\eta Q)\right\} \\ &&
	+\fr{\sigma_0 h_0^2}4
	{\rm Tr}(\Lambda Q). \nn
\eea
One
remarkable thing to keep in mind is that the singlet interaction term
(proportional to $c_0$) can not be treated as an operator insertion as one
would naively expect. It is, in fact, going to affect the ultraviolet
singularity structure of the problem (i.e. poles in $\varepsilon$) and,
hence, changes the $\beta$-functions of the theory. This eventually happens
in the limit where $N_{{\rm \max }}$ is sent to infinity. This is done in
such a way that the quantity 
$W\! =\! T\cdot N_{{\rm \max }}$ remains finite, and
this corresponds physically to to having a finite `effective'
bandwidth $W$.

Next we insert the background field $T_0$ in all but the last term of the
action. The result can be written as
\be
\label{ST_0}
	S[ Q,T_0] =S_k[T_0\inv QT_0] +S_i[T_0\inv QT_0]
	+S_f[T_0\inv QT_0] +\fr{\sigma _0h_0^2}4{\rm Tr}(\Lambda Q), 
\ee
where
\begin{eqnarray}
S_{k}[T_0\inv QT_0]  &=&-\fr{\sigma _{0}}{8}{\rm Tr}\left[
\prt_\mu+A_\mu,Q\right]\left[\prt_\mu+ A_\mu,Q\right] , \\
S_{i}[T_0\inv QT_0]  &=&\fr{\pi}{\qb} z_{0}c_{0}\int d^{D}x\sum_{\alpha n}
{\rm tr}\left\{ (\Ian+A_{n}^{\alpha })Q\right\} {\rm tr}\left\{
(\Iamn+A_{-n}^{\alpha })Q\right\} , \\
S_{f}[T_0\inv QT_0]  &=& \fr{4\pi}{\qb} z_{0}\int d^{D}x{\rm tr}\left\{ (\eta
+A_{\eta })Q\right\} .
\end{eqnarray}
Here all $T_0$ dependence is collected into the `potentials'
$A_\mu$, $A_n^\alpha$ and $A_\eta $
\bea
A_\mu =T_0\partial _\mu T_0^{-1} \hskip8mm &
 A_m^\alpha =T_0\left[
{\rm I}_m^\alpha, T_0^{-1}\right] &
\hskip8mm A_\eta =T_0\left[ \eta
,T_0^{-1}\right] . 
\label{'potentials'}
\eea
The effective action $S\eff[T_0]$ is going to be
obtained by treating the `potentials' $A_{\mu }$, $A_{m}^{\alpha }$
and
$A_{\eta }$ as a perturbation about the original action 
(\ref{Sbackground}). It is easy
to see however, that this procedure breaks the 
$U(N)\!\times\! U(N)$ gauge
invariance which means that the final result can not be expressed, generally
speaking, in terms of the local variable $Q_{0}$ (\ref{Q0}). 
In order to retain 
$U(N)\!\times\! U(N)$ gauge invariance it will be necessary to drop all
temperature and
frequency dependence as infrared regulators such that the parameters in the
effective theory only depend on the regulating field $h_{0}^{2}$. We will
come back to this problem at a later stage (section \ref{secfreqintrenorm}).

\subsection{Perturbation expansion}
\label{secperturb}

The theory (\ref{ST_0}) is going to be worked out perturbatively. Write
\be
	Q=\left(\begin{array}{cc}
	\sqrt{1-qq\dagg} & q \\ 
	q\dagg & -\sqrt{1-q\dagg q}
	\end{array}\right), 
\label{qparametr}
\ee
then the theory can be written as an infinite power series in the matrix
fields $q$, $q\dagg$ which contain $N\!\times\! N$ independent complex field
variables. Notice that classically 
(i.e. $Q\! =\!\Lambda $) the effective
action $S[\Lambda,T_{0}]$ is of the same form as the original one
(\ref{Sbackground}) except that the ${\rm Tr}(\Lambda Q)$ term is lacking.

In order to discuss the theory on a quantum level we regroup the various
quantities in the background field action (\ref{ST_0}) according to 

\bea
\label{Sregrouped}
S[Q,T_{0}] &=& S[Q]+S[T_{0}^{-1}\Lambda T_{0}]-\int\! d^{D}x
\left\{ \fr{\sigma
_{0}}{4}\cdot O_{k}(Q,T_{0})-\fr{4\pi}{\qb}z_0\cdot O_{f}(Q,T_{0}) \right.
\\ &&
\left. -\fr{\pi}{\qb}z_0 c_{0}\cdot O_{i}(Q,T_{0})\vphantom{\fr{\sigma_0}{4}}
\right\} \nn
\eea
where $O_{k}=O_{k}^{(1)}+O_{k}^{(2),1}+O_{k}^{(2),2}\;\;\;\;$ ; 
$\;\;\;\; O_{i}=O_{i}^{(1),1}+O_{i}^{(1),2}+O_{i}^{(2),1}+O_{i}^{(2),2}$. 
\newline
The eight different operators $O$  are listed in Table 1.

\vskip0.5cm

\[
\begin{tabular}{|rl|l|rl|}
\hline
&  &  &  &  \\ 
$O_{k}^{(1)}=$ & $2{\rm tr}(A_{\mu }\delta Q\partial _{\mu }\delta Q)$ &  & $
O_{i}^{(1),1}=$ & $2\sum_{m,\alpha }{\rm tr}({\rm I}_{-m}^{\alpha}Q_{0})
{\rm tr}
({\rm I}_{m}^{\alpha }\delta Q)$ \\ 
&  &  &  &  \\ \hline
&  &  &  &  \\ 
$O_{k}^{(2),1}=$ & $2{\rm tr}(\delta QA_{\mu }\Lambda A_{\mu })$ &  & $%
O_{i}^{(1),2}=$ & $2\sum_{m,\alpha }{\rm tr}({\rm I}_{-m}^{\alpha }\delta Q){\rm tr}%
(A_{m}^{\alpha }\delta Q)$ \\ 
&  &  &  &  \\ \hline
&  &  &  &  \\ 
$O_{k}^{(2),2}=$ & ${\rm tr}(A_{\mu }\delta QA_{\mu }\delta Q)$ &  & $%
O_{i}^{(2),1}=$ & $2\sum_{m,\alpha }{\rm tr}({\rm I}_{-m}^{\alpha }Q_{0}){\rm tr}%
(A_{m}^{\alpha }\delta Q)$ \\ 
&  &  &  &  \\ \hline
&  &  &  &  \\ 
$O_{f}=$ & ${\rm tr}(A_{\eta }\delta Q)$ &  & $O_{i}^{(2),2}=$ & $%
\sum_{m,\alpha }{\rm tr}(A_{-m}^{\alpha }\delta Q){\rm tr}(A_{m}^{\alpha
}\delta Q)$ \\ 
&  &  &  &  \\ \hline
\end{tabular}
\]
Table 1: Classification of terms in $\qd Q\!=\!Q\!-\!\qL$ that
contribute to the background field action; see text.

\vskip0.5cm

The superscript (in brackets) of $O$ shows the lowest order of
$\delta T_{0}\! =\! T_{0}\! -\! 1$ which appears in the decomposition of the
expression in powers of $\delta T_{0}$, while the other one (if present)
just indicates the enumeration of such a contribution.
The propagator for the $q$-field can be read from the quadratic part of the
action (\ref{Sbackground}) and has the form
\be
\left< q_{n_{1}n_{2}}^{\alpha \beta }(p)\;[q\dagg]_{n_{4}n_{3}}^{\delta
\gamma }(-p)\vphantom{H^H}\right>=
\fr{4}{\sigma _{0}}\delta ^{\alpha \gamma }\delta ^{\beta
\delta }
\qd_{n_1-n_2,n_3-n_4}
D_{p}(n_{12})
\left\{ \delta _{n_{1}n_{3}}+\delta ^{\alpha \beta }\kappa
^{2}z_0 c_{0}D_{p}^{c}(n_{12})\right\} ,  \label{23}
\ee
\bea
D_{p}(n_{12})=\frac{1}{p^{2}+h_{0}^{2}+\kappa ^{2}z_0 n_{12}} 
	\hskip0.5cm &;& \hskip0.5cm
D_{p}^{c}(n_{12})=\frac{1}{p^{2}+h_{0}^{2}+(1-c_{0})\kappa ^{2}z_0 n_{12}}
\nn\\
	\qk^2:=\frac{8\pi}{\qb\qs_0} \hskip0.5cm &;& \hskip0.5cm
	n_{12}:=n_1-n_2. \nn
\eea
Here and in what follows we use the following convention: Matsubara
indices with odd subscripts ($n_{1},n_{3},...$) run only over
nonnegative values, while those with even subscripts ($n_{2},n_{4},...$)
run only over negative values. This choice is made in order to
incorporate the structure (\ref{qparametr}), where it is clear that
the first and second index of $q$ have to be nonnegative and negative
respectively. (And vice versa for $q\dagg$).
Often we use the abbreviation 
\be
	\qa=1-c_0.
\ee 
In the case of Coulomb interactions we have 
$\qa\!\naar\! 0$.

In the following sections we present the results of a detailed
computation that shows that the final results for $S\eff$ are of the
same form as the original action (\ref{Sbackground}) with, however,
modified `effective' parameters $\qs'$, $z'$ and $c'$ instead of
$\qs_0$, $z_0$ and $c_0$.

\subsection{Conductivity renormalization}
\label{seccondrenorm}

The contributions to the conductivity $\qs'$ have four different
sources. Schematically, they come from the contractions in the
following terms: 
$\langle O_{k}^{(2),1}\rangle $,
$\langle[O_k^{(1)}]^2\rangle$,
$\langle [O_i^{(1),2}]^2\rangle$  and
$\langle O_k^{(1)}O_i^{(1),2}\rangle$,
with the quantities $O_k$ and $O_i$ defined in Table~1.
The last two of these contributions involve a small momentum expansion
in the background field $T_0$. In Table~2 we present the various
contributions to the pole terms in $1/\qe$ obtained by working in
$D\! =\! 2+2\qe$ dimensions and using dimensional regularization.

\vskip1cm
\[
\begin{tabular}{|l|r|c|c|c|}
\hline
\  & \  & \  & \  & \  \\ 
\  & $\langle O_{k}^{(2),1}\rangle $ & $-{\displaystyle \fr{1}{2}
}\langle[ O_{k}^{(1)}]^2\rangle $ & $-{
\displaystyle \fr{1}{2}}\langle[ O_{i}^{(1),2}]
^{2}\rangle $ & $-\langle O_{k}^{(1)}O_{i}^{(1),2}\rangle $
\\ 
\  & \  & \  & \  & \  \\ \hline
\  & \  & \  & \  & \  \\ 
${\rm tr}(A_{\mu }^{+-}A_{\mu }^{-+}+A_{\mu }^{-+}A_{\mu }^{+-})$ & $\ln
\alpha $ & $0$ & $-(2+{\displaystyle \fr{1+\alpha }{1-\alpha }}\ln \alpha )
$ & $0$ \\ 
\  & \  & \  & \  & \  \\ \hline
\  & \  & \  & \  & \  \\ 
${\rm tr}(A_{\mu }^{++}A_{\mu }^{++}+A_{\mu }^{--}A_{\mu }^{--})$ & $-\ln
\alpha $ & $-2(1+{\displaystyle \fr{\alpha }{1-\alpha }}\ln \alpha )$ & $%
-(2+{\displaystyle \fr{1+\alpha }{1-\alpha }}\ln \alpha )$ & $2(2+{%
\displaystyle \fr{1+\alpha }{1-\alpha }}\ln \alpha )$ \\ 
\  & \  & \  & \  & \  \\ \hline
\end{tabular}
\]
Table 2: Contributions from different sources (top row) to the 
$1/\qe$ pole terms in the background field action (left column); see text.

The most important terms generated by the background field procedure
are 
$(A_\mu)^{\qa\qb}_{n_1 n_2}(A_\mu)^{\qb\qa}_{n_2 n_1}$,
which are indicated by
$\tr A_\mu^{+-}A_\mu^{-+}$ in Table~2, the
$(A_\mu)^{\qa\qb}_{n_1 n_3}(A_\mu)^{\qb\qa}_{n_3 n_1}$,
which are denoted as 
$\tr A_\mu^{++}A_\mu^{++}$
and finally those obtained by interchanging the positive and negative
Matsubara frequency indices. Notice that the first terms can be
written in $\qs$-model form
\be
	\tr\left(A_\mu^{+-}A_\mu^{-+}+A_\mu^{-+}A_\mu^{+-}
	\right)=
	-\kwart\tr(\nabla Q_0)^2.
\ee
The terms on the second line of Table~2 can not be written in terms of
the quantity $Q_0$ as they break the local 
$U(N_{\rm max})\!\times\! U(N_{\rm max})$
gauge invariance. The various contributions in this case sum up to
zero, however, such that the final theory retains the
$U(N_{\rm max})\!\times\! U(N_{\rm max})$
symmetry.
The final result for the effective parameter $\qs'$ becomes
\be
	\qs'=\qs_0\left(1+\fr{4h_0^{2\qe}}{\qs_0\qe}\qO_D
	[1+\fr{\qa}{1-\qa}\ln\qa]\right)
\label{sigprime}
\ee
where 
$\qO_D \! =\!\fr{S_D}{2(2\pi)^D}$ 
with $S_d$ the surface of the unit sphere in
$d$ dimensions. 
Notice that (\ref{sigprime}) behaves smoothly as $\qa$
approaches zero (Coulomb case) whereas the individual contributions
listed in Table~2 do not. On the other hand, (\ref{sigprime}) reduces
to the well known free particle case as $\qa\!\naar\! 1$.
In what follows, it will be convenient to introduce the inverse conductivity $t_0$,
\be
	t_0=4\qO_D/\qs_0.
\ee
We can remove the pole term in $\qe$ from (\ref{sigprime}) by defining
a renormalized theory
$t_0\! =\!\mu^{-2\qe}tZ_1$
as usual, leading to a finite expression in $\qe$.
From (\ref{sigprime}) we extract
\be
	Z_1=1+\fr{t}{\qe}[1+\fr{\qa}{1-\qa}\ln\qa]
\ee
yielding
\be
	\fr{1}{4\qO_D}\qs'
	=t\inv\left(1+t\ln(\mu^2 h_0^2)[1+\fr{\qa}{1-\qa}\ln\qa]\right).
\ee
Notice that the parameter $h_0$ appears as an infrared cutoff rather
than the frequency or temperature, and here it plays the role of the
(inverse) sample size. In the next section we give an explicit example
indicating the kind of complications one is running into by working
with finite temperature or frequency rather than $h_0$.

We now have all the necessary ingredients to calculate the beta
function for the conductance,
\be
	\qb=\frac{dt}{d\ln\mu}
	=2\qe t-2t^2[1+\fr{\qa}{1-\qa}\ln\qa].
\label{betaalpha}
\ee

\subsection{Frequency and interaction renormalization}
\label{secfreqintrenorm}

The singular contributions corresponding to the terms quadratic in
$A_m$ and $A_\qh$ (\ref{'potentials'}) in $S\eff$ originate from many
sources. They are contained in the following contractions:
$\langle O_{f}\rangle $, 
$\langle O_{i}^{(1),2}\rangle $, 
$\langle O_{i}^{(2),1}\rangle $,
$\langle O_{i}^{(2),2}\rangle $, 
$-\half\langle[O_{i}^{(1),2}]^2\rangle $ and 
$-\langle O_{i}^{(1),1}O_{i}^{(1),2}\rangle$,
while all other possible sources give answers which are finite in
$\qe$. Each of the individual contractions gives a divergent answer as
$\qa$ approaches zero, just as in the computation of $\qs'$. However,
numerous cancelations of these divergencies take place in the total
sum.
The resulting expression for $S_{\rm eff}[T_0]$ is given by
\bea
        && -\fr{\pi}{\qb}z_0\left\langle
        4O_f +2c_{0}O_i ^{(1),2}
        +2c_{0}O_i ^{(2),1}+c_{0}O_i ^{(2),2}-
        \fr{\pi}{\qb}z_0 c_{0}^{2}[ O_i ^{(1),2}] ^{2}-\fr{2\pi}{\qb}z_0
        c_{0}^{2}\;O_i ^{(1),1}O_i ^{(1),2}\right\rangle
        \nn\\ && \longrightarrow
        \qG_0+\sum_{\qa\qb\qg}{\sum_{m n_1 n_2}\!\!}' \;
        \qG_1(m,n_1,n_2)(T_0{\rm I}^\qa_m T_0\inv)^{\qb\qg}_{n_2 n_1}
        (T_0{\rm I}^\qa_{-m}T_0\inv)^{\qg\qb}_{n_1 n_2}
\label{IAIA}
\eea
with $\qG_1$ defined as
\be
        \qG_1(m,n_1,n_2)=-\qk^2 z_0 c_0\int_p D_p(n_{12})
        \left[1+\qk^2 z_0 c_0 |m| D_p^c(|m|)\right],
\label{defGamma1}
\ee
and where the prime on the frequency summation indicates a restriction
on the frequency range,
$m,n_1,n_2\!\in\!\{-2N_{\rm max}\!+\!1,\cdots,2N_{\rm max}\!-\!1\}$
(see~\cite{3A}).
In fact, the sum over frequencies outside of this range cancels the
contribution from $\langle O_f\rangle$.

In (\ref{IAIA}) we have included a constant $\qG_0$ which is such that
it cancels the constant that appears in the $\qG_1$ term. The singular
contribution is obtained by replacing the square bracket in
(\ref{defGamma1}) by unity, and we get simply
\be
        \int_p D_p(n_{12})=-\fr{\qO_D}{\qe}(h_0^2+\qk^2 z_0 n_{12})^\qe
        \approx -\fr{\qO_D}{\qe}-\qO_D\ln(h_0^2+\qk^2 z_0 n_{12}).
\label{sing}
\ee
We mention two important aspects of this result. First, the pole term in
$\qe$ is independent of the frequency indices $n_1,n_2$ and this implies
that the result (\ref{IAIA}) becomes 
$U(N_{\rm max})\!\times\! U(N_{\rm max})$ invariant, i.e. it can be expressed
in the quantity $Q_0$ (see below). Notice that 
$U(N_{\rm max})\!\times\! U(N_{\rm max})$ gauge invariance is
automatically
obtained by putting the ratio $\qk^2 z_0 n_{12}/h_0^2$ equal to zero, i.e. by
dropping the temperature and frequency as infrared regulator in the final
answer. This, however, is just a different way of saying that the
background field procedure, along with any other arbitrary RG program,
can never provide information on the dynamics of the 
problem, such as the AC conductivity.

Secondly, we evaluate the result (\ref{IAIA}) as follows.
\bea
\label{Q0form}
        S_{\rm eff}[T_0] &=&
        \qG_0-\qk^2 z_0 c_0 \fr{h_0^{2\qe}}{\qe}\qO_D\sum_{\qa\qb\qg}
        {\sum_{m n_1 n_2}\!\!}' \;
	(T_0 {\rm I}^\qa_m T_0\inv)^{\qb\qg}_{n_2 n_1}
        (T_0 {\rm I}^\qa_{-m}T_0\inv)^{\qg\qb}_{n_1 n_2} \\
        &=&
        \qG_0+\qk^2 z_0 c_0\fr{h_0^{2\qe}}{8\qe}\qO_D{\sum_{m\qa}}'
        \tr[T_0 {\rm I}^\qa_m T_0\inv,\qL][T_0 {\rm
        I}^\qa_{-m}T_0\inv,\qL] \nn\\
        &=& 
        \qG_0+\qk^2 z_0 c_0\fr{h_0^{2\qe}}{8\qe}\qO_D{\sum_{m\qa}}'
        \tr [{\rm I}^\qa_m,Q_0][{\rm I}^\qa_{-m},Q_0] \nn
\eea
which has precisely the ${\cal F}$-invariant form in $Q_0$ 
given in (\ref{SF}). 
Hence, the significance of working with `small' background fields $T_0$
(relative to the field variables $Q$) is now well recognized and the
renormalization procedure for the interacting electron gas involves not
only the usual rescaling of momenta, it necessarily involves also a
rescaling of the time or frequency variable or, rather, of the `effective'
bandwidth $W\! =\! TN_{\rm max}$.

Next, from (\ref{Q0form}) we can read off the numerical value for the
constant $\qG_0$ which is given by
\be
	\qG_0=\fr{3}{2}\qk^2 z_0 c_0\fr{h_0^{2\qe}}{\qe}\qO_D N_r \sum_m |m|.
\ee
It can be shown (Section \ref{seconeloop}) 
that $\qG_0$ is precisely canceled by the
one-loop result for the free energy (${\cal F}_1$) which still appears
in the definition of $S\eff$. Hence the expression in $Q_0$
(\ref{Q0form}) is an eigenoperator of the renormalization procedure, and
this is in sharp contrast to what one is used to in the conventional
$\qs$-model theory applied to the free electron problem \cite{a22}. 
Operators
bilinear in $Q$ as well as higher order operators in this case describe
density fluctuations which become anomalous (multifractal) as one
approaches the critical point (mobility edge) in $2\! +\! 2\qe$ dimensions. The
physics of the interacting electron gas appears to be quite different in
this respect and, as will be shown in later sections, the theory is much
closer to the familiar one describing the Heisenberg ferromagnet. In
particular, the interacting system is characterized by a conventional
order parameter, as well as a (Coulomb) gap in the quasiparticle density
of states which enters the expression for the specific heat.

We summarize the results of this section by giving the complete form for
the effective background field action (including constants)
\bea
	S\eff[Q_0] &=& -\fr{\qs'}{8}\Tr(\nabla Q_0)^2
	+\fr{\pi}{2\qb}z'{\sum_{\qa n}}'\Tr[\Ian,Q_0][\Iamn ,Q_0]
	\nn\\ &&
	+\fr{\pi}{\qb}\qa'z'\int_x{\sum_{n\qa}}\tr \Ian Q_0\;
	\tr \Iamn Q_0
\label{SQ0prime}
\eea
where the effective parameters $\qs'$, 
$c' \! =\! 1 \! -\!\qa'$ and $z'$ are given by
(\ref{sigprime}) and 
\bea
\label{zprime}
	z' &=& z_0(1+\fr{2}{\qs_0}c_0\qO_D\fr{h_0^{2\qe}}{\qe})=
	z_0(1+\fr{h_0^{2\qe}}{2\qe}c_0 t_0)  \\
	\qa'z' &=& (1-c')z'=(1-c_0)z_0=\qa_0 z_0 \nn.
\eea
Introducing renormalization constants for the amplitudes,
$z_0 \! =\! \mu^{2\qe}zZ_2$ and 
$\qa_0 \! =\!\qa Z_\qa$, we then extract from (\ref{zprime})
\bea
	Z_\qa=Z_2\inv \hskip0.5cm &;& \hskip0.5cm Z_2=1-\fr{t}{2\qe}c
\eea
such that the renormalization of the $\qa$ is obtained as 
\be
	\fr{d\qa}{d\ln\mu}=-t\qa(1-\qa)
\label{renormalpha}
\ee
and the anomalous dimension of the amplitude $z$ is expressed in terms of
the $\qg$-function
\be
	\qg=-\fr{d(\mu^{2\qe}z)}{d\ln\mu}=\fr{d\ln Z_2}{d\ln\mu}
	=-t(1-\qa).
\label{gammaz}
\ee
Equation (\ref{renormalpha}) implies that ${\cal F}$-invariance of the
theory ($\qa\! =\! 0$) is retained by the renormalization group. The theory
is unstable with respect to the symmetry breaking $\qa$ which plays a
role analogous to the `range' of the electron-electron interactions
\cite{PruiskenBaranov}. From the $\qb$-function (\ref{betaalpha}) and 
(\ref{gammaz}) we conclude that the Coulomb
interaction problem ($\qa\! =\! 0$) has a fixed point 
$t_c\!\propto\! \qe$ in
$2\! +\! 2\qe$ dimensions which separates a metallic phase 
($t\! <\! t_c$) from an
insulating phase ($t\! >\! t_c$). In two spatial dimensions, the metallic
phase disappears altogether and this, then, leads to the familiar
complications as far as the quantum Hall effect is concerned 
\cite{PruiskenBaranov}.

\ns{Linear response}
\label{seclinresp}

In the previous section we have seen that within the background field
approach, the singlet interaction term contributes in a peculiar fashion
to the conductivity renormalization. In this section we present the
results of a detailed computation of $\qs'$ by considering the linear
response to a vector potential insertion. If one views the insertion of a
vector potential as the result of a (`large') background field rotation
$W$, then it is clear that this special type of background field leaves
the Finkelstein action invariant, the symmetry being broken only by
$\qa\!\neq\! 0$ for $h_0^2\! =\! 0$.

In contrast to the background field procedure, we can now expect that the
limit $h_0\!\naar\! 0$ can be taken in the end such that the frequency and $T$
dependence of the $\qs'$ can be computed.
We stress that it is a priori completely un-obvious that the two different
ways of computing $\qs'$ should yield identical renormalization group
results in the end. In fact, the different methods correspond to two
completely different manners of handling the small frequency, long
wavelength excitations of the system. In a previous paper \cite{3A} we have
shown that these different methods are formally related by the
$U(N_{\rm max})\!\times\! U(N_{\rm max})$
gauge invariance of the theory.
Hence, the results of this section should be considered as an important
demonstration of the significance of the ${\cal F}$-invariance present in the
problem. 

We start out by replacing the derivatives in (\ref{Sbackground}) by
covariant derivatives
\be
	\prt_j Q\naar [\prt_j-i\hat A_j,Q].
\ee
The `effective' action for the vector potential $\vec A$ is defined by
\be
	e^{S\eff[\vec A]}=\pathint{[Q]}e^{S[Q,\vec A]}
\ee
which is generally given as
\be
	S\eff[\vec A]=\int_x\sum_{\qa,n>0} \qs'(n) n
	[\vec A^\qa_n]^* \cdot\vec A^\qa_n.
\ee
The unknown quantity $\qs'(n)$ can be formally expressed in terms of
correlations of the $Q$-fields as follows
\bea
\label{sigprimecorr}
	\qs'(n) &=& -\fr{\qs_0}{4n}\tr\left\langle\vphantom{\int}
	[\Ian,Q][\Iamn ,Q]\right\rangle\\
	&&
	+\fr{\qs_0^2}{16nD}\int_{x'}\left\langle\vphantom{\int}
	\left\{\tr[\Ian Q]\nabla Q(x)\right\}\cdot
	\left\{\tr[\Iamn  Q]\nabla Q(x')\right\}
	\right\rangle\nn
\eea
where the expectation is with respect to the theory of
(\ref{Sbackground}). Here we take a single fixed replica
channel. At a classical level ($Q\! =\!\qL$), equation (\ref{sigprimecorr})
gives us $\qs' \! =\!\qs_0$ as it should. The quantum fluctuations of the
$Q$-field give rise to various contributions which are individually
singular as $\qa$ approaches zero. A detailed computation gives the
following answer
\bea
	\qs'(n) &=& \qs_0-\fr{16\qk^4 c_0}{D}\fr{2\pi}{\qb}z_0^2
	\sum_{m=1}^\infty
	m\int_p p^2 D_p(m+n)D_p(m+2n)D_p^c(m)\times\nn\\ &&\times
	[D_p(m)+\fr{1-c_0}{2}D_p^c(m+n)].
\eea
The sum and integral can be performed by keeping one of the regulators $h_0^2$,
temperature or frequency 
$\qo\! =\!\fr{2\pi}{\qb}n$ fixed and putting the
other two to zero.
The general result involves complex expressions containing dilogs and
$\psi$-functions. Here we limit the discussion to the most interesting
case of the Coulomb interaction 
($\qa\! =\! 1 \! -\! c_0 \! =\! 0$). We find to lowest order
in $\qe$
\bea
	\qs'(T) &=& \qs_0-4\qO_D(-\fr{1}{\qe}-\ln(\fr{8\pi}{\qb\qs_0}z_0)
	+2+\qg) \\
	\qs'(\qo) &=& \qs_0-4\qO_D(-\fr{1}{\qe}-\ln\fr{z_0\qo}{\qs_0}
	+\fr{3}{2}-2\ln 2) \\
	\qs'(h_0^2) &=& \qs_0-4\qO_D(-\fr{1}{\qe}-\ln h_0^2).
\label{sigprimeh0}
\eea
Here, $\qg\!\approx\! 0.5772\ldots$ denotes Euler's constant. As before, the
numerical factor $\qO_D$ can be absorbed in a redefinition of the
parameters $\qs'$, $\qs_0$. Notice that (\ref{sigprimeh0}) is precisely the
result we previously obtained by employing the background field procedure.
A good approximation to the general case is obtained by replacing $\ln h_0^2$
in (\ref{sigprimeh0}) by the following expression with appropriately
chosen constants $a$ and $b$
\be
	\ln h_0^2\naar \ln[h_0^2+a\fr{z_0\qo}{\qs_0}+b\fr{z_0 T}{\qs_0}].
\ee
This expression indicates that the square of the phase breaking length
($L_\qf^2$) due to the different mechanisms of sample size, frequency and
temperature add up in parallel. We have convinced ourselves of this result
by performing numerical analyses.

\ns{Free Energy}
\label{secfree_energy}
\subsection{Introduction}

In this section we build upon an important result obtained from the
background field procedure. In particular the construction of the
eigenoperators (which includes the constants in the quantum fluctuations
$q,q\dagg$) can be exploited in a very effective fashion by extracting
higher loop renormalization results from a computation of the free
energy {\cite{foot1}},
in complete analogy with what has been done for ordinary
$\qs$-models \cite{BHZJ}.
However, the computation of the free energy in our case
provides additional information not only on the theory in strong coupling,
it also gives rise to the identification of a new physical quantity in the
problem, namely the quasiparticle density of states entering the specific
heat.

Since the apparatus of the renormalization group can only be put to
work once the singularity structure of the free theory is known, we shall
proceed by first reporting the results of a detailed computation of the
free energy to two-loop order. These results indicate that
the quantum theory retains the tree level structure of the free energy, a
result which is obtained by putting $Q_0\! =\!\qL$ in our background field
functional (\ref{SQ0prime}). More specifically, write
\be
	S\eff[Q_0=\qL]=-2V_D N_r {\sum_{n>0}}'z'\qo_n
\label{Sclass}
\ee
where the prime on the summation sign now merely indicates that the limit
of small frequencies $\qo_n\!=\!\fr{2\pi}{\qb}n$ is of interest only. 
The $V_D$ stands for the volume of the $D$-dimensional system.

For reasons which have been well
explained in sections \ref{secbackground} 
and \ref{seclinresp}, the background field procedure only
provides information on the quantity $z'$ in the case where the $h_0^2$
provides the dominant infrared regularization. Since ${\cal F}$-invariant
quantities like the free energy are, in principle, free of such a
constraint, it is important to know whether and how the regulator $h_0^2$
can be put equal to zero. Most of the answer to the question is already
provided by the theory in one-loop approximation which will be discussed
in section \ref{seconeloop}. 
In section \ref{sectwoloop} we report the results of the much more
complicated theory at two-loop level. The reader who is not interested in
technical details might skip this section and proceed immediately to
section \ref{secscaling} 
where the scaling implications are analyzed, making use (amongst
other things) of the more familiar theory of the Heisenberg ferromagnet.

\subsection{One-loop theory}
\label{seconeloop}

Let us proceed from the results obtained from the background field
procedure (\ref{SQ0prime}) and start from the renormalizable action
\be
	S[Q] = -\fr{\qs_0}{8}\Tr (\nabla Q)^2
	+\fr{\pi}{2\qb}z_0{\sum_{\qa n}}' \;
	\Tr[\Ian,Q][\Iamn ,Q]
	-\fr{\pi}{\qb}z_0 \qa_0{\sum_{n\qa}}\int_x
	\tr \Ian Q\;\tr \Iamn Q.
\ee
This action can be written in a form where the `large' diagonal components
of $Q$ do not contribute,
\bea
\label{SdeltaQ}
	S[Q] &=& -2z_0 V_D N_r{\sum_{n>0}}' 
	\qo_n-\fr{\qs_0}{8}\Tr (\nabla Q)^2
	+\fr{\pi}{\qb}z_0 c_0{\sum_{n\qa}}\int_x
	\tr \Ian Q\;\tr \Iamn Q \\
	&& +4z_0\fr{\pi}{\qb}\Tr\qh(Q-\qL). \nn 
\eea
The free energy ${\cal F}$ is defined by
\be
	e^{-V_D N_r\cal F}=\pathint{[Q]}e^{S[Q]}
\ee
and can be written as
\be
	{\cal F}={\cal F}_0+{\cal F}_1+{\cal F}_2+\cdots
\ee
where the subscript $i$ on ${\cal F}_i$ indicates the `order' of the loop
expansion. The quantity ${\cal F}_0$ is given by the first term in
(\ref{SdeltaQ}),
\be
	{\cal F}_0=2z_0{\sum_{n>0}}' \qo_n.
\label{F0}
\ee
On the other hand, the ${\cal F}_1$ is readily obtained from the Gaussian
quantum theory and the result is
\bea
	{\cal F}_1 &=& \sum_{n>0}\int_p\ln\frac{p^2+h_0^2+\qa\qk^2 z_0 n}
	{p^2+h_0^2+\qk^2 z_0 n} \nn\\
	&=& 
	-\qO_D\sum_{n>0}\frac{(h_0^2+\qk^2 z_0 n)^{1+\qe}
	-(h_0^2+\qa\qk^2 z_0 n)^{1+\qe}}{\qe(1+\qe)}.
\label{F1}
\eea
By taking the $\ln T$-derivative of ${\cal F}$ one reproduces the result of
the background field procedure (\ref{Sclass}),
\be
	\frac{\prt{\cal F}}{\prt\ln T}=2{\sum_{n>0}}'z_0\qo_n
	\left\{1+\frac{2\qO_D}{\qs_0}\frac{(h_0^2+\qk^2 z_0 n)^\qe
	-\qa(h_0^2+\qa\qk^2 z_0 n)^\qe}{\qe}\right\}.
\label{dFdlnT}
\ee
By neglecting the $\qo_n$ terms relative to the infrared regulator
$h_0^2$, one obtains 
\be
	\frac{\prt{\cal F}}{\prt\ln T}=2z'{\sum_{n>0}}'\qo_n
\ee
with $z'$ precisely given by the background field result of (\ref{Sclass}).
Next we consider (\ref{dFdlnT}) in the limit 
$h_0^2\!\naar\! 0$. In this case
(\ref{dFdlnT}) defines a $\qo_n$-dependent effective parameter
$z'(\qo_n)$,
\be
	\frac{\prt{\cal F}}{\prt\ln T}=2{\sum_{n>0}}'z'(\qo_n)\qo_n
\ee
where
\be
	z'(\qo_n)=z_0\left\{1+\fr{2\qO_D}{\qs_0}(\fr{4z_0\qo_n}{\qs_0})^\qe
	\fr{1-\qa^{1+\qe}}{\qe}\right\}.
\ee


\subsection{Two-loop theory}
\label{sectwoloop}

Treating the nonquadratic part of the Hamiltonian as a perturbation, the
expression for the two-loops contribution to the free energy can be
presented in the form

\begin{equation}
	{\cal F}_2=-\left\langle S_{0}^{(4)}+S_{{\rm int}}^{(4)}+\frac{1}{2} 
	\left( S_{{\rm int}}^{(3)}\right) ^{2}\right\rangle ,  \label{2.1}
\end{equation}
where the superscript between brackets denotes the order in
$q,q\dagg$.
The $S_0$ stands for the `free' action without the interaction term,
and its fourth order part is given by
\begin{eqnarray}
	S_{0}^{(4)} &=&\frac{\qs_0}{32}\int_p
	\sum q_{n_{1}n_{2}}^{\alpha \beta }(p_{1}) 
	[q_{n_{3}n_{2}}^{\gamma \beta }(p_{2})]^*q_{n_{3}n_{4}}^{\gamma
	\delta }(p_{3})[q_{n_{1}n_{4}}^{\alpha \delta }(p_{4})]^*
	\cdot \delta (p_{1}+p_{2}+p_{3}+p_{4})  \nn
	\times\\ &&\times
	\left\{\vphantom{H^{H^H}}
	 (p_{1}{-p}_{2})\cdot(p_{3}{-p}_{4})+(p_{1} 
	{-p}_{4})\cdot(p_{3}{-p}_{2})
	+\qk^2 z_0(n_{12}+n_{34})+2h_{0}^{2}\right\}.
\end{eqnarray}
The fourth and third order part of the interaction
term are given by
\begin{equation}
	S_{{\rm int}}^{(4)}=\fr{\pi}{4\qb}z_0 c_0
	\sum_\qa\left\{ {\rm tr}\left( {\rm I}_{0}^{\alpha }
	\left[ q,q\dagg\right] \right) \right\} ^{2}
	+\fr{\pi}{2\qb}z_0 c_0\sum_{\alpha ,n>0}{\rm tr}
	\left( {\rm I}_n^{\alpha }\left[q,q\dagg\right] \right) 
	{\rm tr}\left( {\rm I}_{-n}^{\alpha }\left[q,q\dagg\right]\right)
\label{2.3}
\end{equation}
and
\begin{equation}
	S_{{\rm int}}^{(3)}=-\fr{\pi}{\qb}z_0 c_0\sum_{\alpha ,n>0}
	\left\{ \tr\Ian q\dagg
	{\rm tr} \left({\rm I}_{-n}^{\alpha}
	\left[ q,q\dagg\right]\right) 
	+{\rm tr} \left({\rm I}_{n}^{\alpha}\left[q,q\dagg\right]\right) 
	\tr \Iamn q \right\}.
\label{2.4}
\end{equation}
Performing the contractions in (\ref{2.1}) is straightforward but
laborious. In the end we get an expression for ${\cal F}_2$
in terms of the basic propagators $D$ and $D^c$
\begin{eqnarray}
{\cal F}_2 &=&\fr{1}{\qs_0}\qk^2 z_0 c_0\sum_{s>0}s\cdot
\int_{p,q}D_{q}(s)D_{p}(s)  \label{2.5.1} \\
&&+\fr{2}{\qs_0}(\qk^2 z_0 c_0)^2\sum_{m,n>0}\min
(m,n)\int_{p,q}\left\{ -D_{p}(n)D_{p}^{c}(n)D_{q 
}(m+n)\right.  \nn\\
&&+D_{q}^{c}(m)D_{p}(n)D_{p}^{c}(n)  \nn\\
&&-D_{q}^{c}(m)D_{p}^{c}(n)D_{{\bf p+q}}(m+n)  \nn\\
&&\left. -\qk^2 z_0 c_0\cdot m\cdot D_{q}^{c}(m)D_{p}(n)D_{p 
}^{c}(n)D_{{\bf p+q}}(m+n)\right\} .  \nn
\end{eqnarray}
The overall factor $\min(n,m)$ in front of all the terms with more
than two propagators can
be roughly understood as follows. The summation 
$m\!\! >\!\! 0$ always enters
the stage when a redefinition occurs of the summation variable 
$n_2$ on $q_{n_1 n_2}$ to a new summation variable 
$m\! =\! n_1\! -\! n_2\! >\! 0$. 
This shift leaves behind a constraint on the range of $n_1$, namely
$n_1\! <\! m$. The tracing with ${\rm I}_n$ in (\ref{2.3}) and (\ref{2.4})
induces a shift and a constraint of the type $n_1\! <\! n$ (or 
$n_1\! +\! n\!\geq\! m$ or something similar). 
In every contribution a free sum over $n_1$ finally occurs,
yielding $\min(n,m)$ due to the constraints imposed on the summation interval.

We wish to present the result of (\ref{2.5.1}) in the
form (\ref{F0}, \ref{F1}), 
i.e. as a frequency sum of corrections to $z_0$.
This means we have to perform two momentum integrals
and one frequency sum, while keeping one frequency fixed. 
This can be done in the following way:
\begin{eqnarray*}
	\sum_{m,n>0}\min (m,n)\cdot f(m,n) 
	&=&\sum_{m=1}^{\infty }m\sum_{n=m+1}^{\infty }f(m,n)
	+\sum_{n=1}^{\infty}n\sum_{m=n+1}^{\infty }f(m,n) \\
	&=&\sum_{s=1}^{\infty }s\left\{ \sum_{l=s+1}^{\infty}
	f(s,l)+\sum_{l=s+1}^{\infty }f(l,s)\right\}.
\end{eqnarray*}
Thus (\ref{2.5.1}) can be rewritten in the form
\begin{eqnarray}
	{\cal F}_2 &=&\fr{1}{\qs_0}\qk^2 z_0 c_0\sum_{s>0}s\cdot
	\int_{p,q}D_{q}(s)D_{p}(s)  \label{2.6.1} \\
	&&+\fr{2}{\qs_0}(\qk^2 z_0 c_0)^2
	\sum_{s>0}\sum_{l>s}\int_{p,q}*\left\{ -D_{q}(s)D_{q}^{c}(s)
	D_{p}(s+l)\right.   
\label{2.6.2} 
	\\ &&
	-D_{q}(s+l)D_{p}(l)D_{p}^{c}(l)  
\label{2.6.3} 
	\\ &&
	+D_{q}^{c}(s)D_{p}(l)D_{p}^{c}(l)  
\label{2.6.4} 
	\\&&
	+D_{q}(s)D_{q}^{c}(s)D_{p}^{c}(l)  
\label{2.6.5} 
	\\&&
	-2D_{q}^{c}(s)D_{p}^{c}(l)D_{{\bf p+q}}(s+l)  
\label{2.6.6} 
	\\&&
	-\qk^2 z_0 c_0\cdot D_{q}(s)D_{q}^{c}(s)\cdot l\cdot D_{p}^{c}(l)
	D_{{\bf p+q}}(s+l)  
\label{2.6.7} 
	\\&&
	\left. -\qk^2 z_0 c_0\cdot s\cdot D_{q}^{c}(s)D_{p}(l)D_{p}^{c}(l)
	D_{{\bf p+q}}(s+l)\right\} .  
\label{2.6.8}
\end{eqnarray}
Calculations are now straightforward but cumbersome. Some details are
presented in the Appendix. The final result for the poles in $\varepsilon $ in
two loop contribution is given by

\bea
{\cal F}_2 &=& \fr{2}{\qs_0}\qk^2 z_0\sum_{s>0}s\cdot
h_{0}^{4\varepsilon }\Omega _{d}^{2}\left\{
-\frac{(1+h_s^2/h_0^2)^{\varepsilon }-1}{\varepsilon \cdot h_s^2/h_0^2}
\cdot \frac{2+\ln \alpha 
}{\varepsilon }-\frac{\ln (1+h_s^2/h_0^2)}{\varepsilon }\right.\nn\\
&& \left.
+\frac{1}{
2\varepsilon ^{2}}+\frac{\pi ^{2}}{2\varepsilon }+{\cal O}(\varepsilon ^{0})\right\} ,
\label{2.7}
\eea
where we have introduced $h_s^2=\qk^2 z_0 s$ and
put $\alpha\! =\! 0$ whenever possible. It should be mentioned that in
obtaining (\ref{2.7}) drastic cancelations of many singular terms in $\alpha $
take place (see Appendix).
Combining (\ref{F1}) and (\ref{2.7}), one gets
\begin{equation}
	{\cal F}_1+{\cal F}_2=\Omega _{d}\cdot \frac{1}{\varepsilon }
	\frac{({h'}^{2}+{h_s'}^2)^{1+\varepsilon }
	-({h'}^{2})^{1+\varepsilon }}{1+\varepsilon }
	+\frac{1}{\qs'}\sum_{s>0}{h_s'}^2  
	\cdot h_{0}^{4\varepsilon }\Omega _{d}^{2}
	\left\{ \frac{1}{\varepsilon ^{2}}+\frac{\pi^{2}}{\varepsilon}
	\right\},
\label{2.8}
\end{equation}
where we make use of the one-loop renormalization of $h_{0},\qs_0$ and $z_0$:
\begin{eqnarray}
	t_0 &\rightarrow &t^{\prime }=t_0\cdot 
	\left\{ 1-\frac{h_0^{2\qe}}{\varepsilon }t_0\right\}  
\label{2.9.1} 
	\\
	z_0 &\rightarrow &z^{\prime }=
	z_0\cdot \left\{ 1+\frac{h_0^{2\qe}}{2\qe }t_0\right\}  
\label{2.9.2}
	\\
	h_{0}^{2} &\rightarrow &{h'}^{2}=h_{0}^{2}\cdot 
	\left\{1-[2+\ln\alpha]\frac{h_{0}^{2\qe}}{\qe}t_0\right\}.
\label{2.9.3}
\end{eqnarray}
Equations (\ref{2.9.1})-(\ref{2.9.3}) imply the following
renormalization of the quantity $h_s^2\propto t_0 z_0$
\be
	h_s^2\naar {h_s'}^2=h_s^2\cdot\left\{1-\frac{h_0^{2\qe}}{2\qe }t_0
	\right\}.
\ee
Let us also present the result for the case when 
$h_{0}^{2}\! =\! 0$, so that
frequency serves as infrared regulator:
\begin{eqnarray}
	{\cal F}_0 &=& \frac{2\qO_D}{t_0}{\sum_{s>0}}' h_s^2 \\
	{\cal F}_1 &=&\frac{2\qO_D}{t_0}{\sum_{s>0}}' h_s^2\cdot
	\frac{h_s^{2\qe}}{\qe(1+\qe)}\frac{t_0}{2}
\label{2.10.1} 
	\\
	{\cal F}_2 &=&-\frac{2\qO_D}{t_0}{\sum_{s>0}}' h_s^2\cdot
	\frac{h_s^{2\qe}}{\qe^2}\frac{t_0^2}{8}
	\left[1-\qe(\frac{\pi^2}{3}+4)+{\cal O}(\qe^2)\right].
\label{2.10.2}
\end{eqnarray}

\ns{Scaling results}
\label{secscaling}

Our renormalization group program of the Finkelstein theory consisted
of two separate parts. First we studied the general background field
procedure in dimensional regularization and we focused primarily on
how the interaction term affects the ultraviolet behaviour of the
theory. This procedure provides two renormalization constants, $Z_1$
and $Z_2$. (From now on we consider the Coulomb case $\qa\! =\! 0$.)

The background field procedure and other, arbitrary, renormalization
group programs \cite{BelitzKirkpatrick} do not provide us, however,
with the full temperature and frequency dependence of physical
observables such as the conductivity. 
This was the subject of the second part in which we
recognized that observables should be constructed out of only those
correlations or `background' fields that leave the interaction term
$S_{\rm F}$ invariant. There are two classes of such correlations: the
free energy itself and the one denoted as `linear response', which
provide us with equilibrium statistical mechanics and transport
theory, respectively. For these quantities we can derive general
scaling results from the renormalization group.

First we summarize the results for the conductivity $\qs_0$ and the
free energy 
${\cal F}_0\! +\!{\cal F}_1\! +\!{\cal F}_2$ as follows
\bea
\label{renormsig}
	\qs'(s) &=& \fr{4\qO_D}{t_0}R_\qs(t_0,h_s) \\
	R_\qs &=& 1+\fr{h_s^{2\qe}}{\qe}t_0\left(1-\qe[\fr{5}{2}-2\ln 2]
	\right). \nn
\eea
The $\ln T$-derivative of the free energy can be written as 
\be
	\fr{\prt{\cal F}}{\prt\ln T}=2\qO_D\sum_{s>0}(h_s')^2/t'=
	2\qO_D\sum_{s>0}\fr{h_s^2}{t_0}M(t_0,h_s)
\ee
where
\be
	M(t_0,h_s)=1+\fr{h_s^{2\qe}}{2\qe}t_0+\fr{h_s^{4\qe}}{\qe^2}t_0^2
	\left(-\fr{1}{8}+\qe\fr{1}{4}[1+\fr{\pi^2}{6}]\right).
\label{M}
\ee
These results are completely analogous to what has been obtained for the
Heisenberg ferromagnet \cite{PruiskenWang}.
The quantity $M$ is the analogue of the magnetization in ferromagnetic
language and we shall next follow up on the analysis of
\cite{PruiskenWang}. Equation (\ref{M}) can be understood in terms of
the effective parameters $h_s'$ and $t'$ with the latter defined by
\bea
	\fr{1}{t'}=\fr{1}{t_0}R(t_0,h_s) \hskip0.5cm &;& \hskip0.5cm
	R(t_0,h_s)=1+\fr{h_s^{2\qe}}{\qe}t_0+{\cal O}(t_0^2).
\label{deft'}
\eea
Notice that the expressions for $\qs'$ and $1/t'$ in (\ref{renormsig}) and
(\ref{deft'}) are not necessarily the same, since they describe
different physics (linear response and equilibrium statistical mechanics
respectively). Nevertheless, both (\ref{renormsig},\ref{M}) and
(\ref{deft'}) can be used to extract the renormalization constants $Z_1$
and $Z_2$ obtained by putting
\bea
	t_0=\mu^{-2\qe}tZ_1 \hskip0.5cm &;& \hskip0.5cm 
	\qO_D\fr{h_s^2}{t_0}=\qo_s z_0=\qo_s \mu^{2\qe}z Z_2
\eea
where $t$, $z$ now stand for the parameters of the renormalized theory.
Following the scheme of minimal subtraction we get
\bea
	Z_1 &=& 1+t/\qe \\
	Z_2 &=& 1-\fr{t}{2\qe}-\fr{t^2}{\qe^2}\left(\fr{1}{8}
	+\qe[\fr{\pi^2}{24}+\fr{3}{4}]\right).
\eea
The renormalization $\qb$- and $\qg$-functions are obtained as usual
\bea
	\qb=\frac{2\qe t}{1+t\fr{d\ln Z_1}{dt}} 
	\hskip0.5cm &;& \hskip0.5cm
	\qg=\qb\frac{d\ln Z_2}{dt}
\eea
yielding
\bea
	\qb=2\qe t-2t^2 \hskip0.5cm &;& \hskip0.5cm
	\qg=-t-t^2(3+\pi^2/6).
\eea
Next we express the free energy and conductivity in terms of the
renormalized parameters
\bea
	\fr{\prt{\cal F}}{\prt\ln T}=2\sum_{s>0} \mu^{2\qe}z\qo_s M 
	\hskip0.5cm &
	\fr{1}{t'}=\mu^{2\qe}\fr{1}{t}R \hskip0.5cm &
	\qs'=\mu^{2\qe}\fr{4\qO_D}{t}R_\qs
\eea
where $M$, $R$ and $R_\qs$ can be written in scaling form following the
method of characteristics
\bea
	M &=& M_0(t)g(\qo_s z\xi^D M_0) \nn\\
	R &=& R_0(t)h(\qo_s z\xi^D M_0) \\
	R_\qs &=& R_0(t)f(\qo_s z\xi^D M_0)\nn
\eea
with $M_0$, $R_0$ and $\xi$ determined from the renormalization group
functions $\qb$ and $\qg$ according to
\bea
	&& [\mu\prt_\mu+\qb\prt_t]\xi(t) = 0 \nn \\
	&& [\qb\prt_t-\qg]M_0(t) = 0 \\
	&& [\qb\prt_t+2\qe-t\inv\qb]R_0(t) = 0. \nn
\eea
More explicitly, in $2\! +\! 2\qe$ dimensions we have for the metallic phase
($t\! <\! t_c$) 
\bea
	\xi &=& \mu\inv t^{1/2\qe}(1-t/t_c)^{-\nu} \nn\\
	R_0 &=& (1-t/t_c)^{2\qe\nu} \\
	M_0 &=& (1-t/t_c)^{\qb_0}\nn
\eea
where $t_c\! =\! \qe\! +\!{\cal O}(\qe^2)$ 
is the critical point ($\qb(t_c)\! =\! 0$)
whereas
\bea
	\nu=-1/\qb'(t_c) \hskip0.5cm &;& \hskip0.5cm \qb_0=-\nu\qg(t_c).
\eea
It is interesting to conclude that the interacting electron gas is much
closer to the physics of the Heisenberg ferromagnet than the free
electron problem which has $\qb_0\! =\! 0$, indicating that the density of
states is nonsingular. The metal/insulator transition for interacting
electrons is characterized by the appearance of a real order
parameter. We will show below (section \ref{secspecheat}) 
that $\qb_0\!\neq\! 0$ in this
case means that the specific heat becomes singular as one approaches
from the metallic side.

Important conceptual quantities of the theory are the various length
scales ($L_\qf$) induced by the phase breaking parameters such as
temperature $T$, frequency $\qo$ and the parameter denoted as $h$. From
the renormalization group we have
\bea
	L_\qf^{-D}(h_0) &=& (h')^2 \nn\\
	L_\qf^{-D}(T) &=& T z M_0 g_T(Tz\xi^D M_0) \\
	L_\qf^{-D}(\qo) &=& \qo zM_0 g_\qo(\qo z\xi^D M_0). \nn
\eea
Finally, we can obtain `equations of state' for the quantities $M$ and
$R$. As shown in \cite{PruiskenWang} to lowest order in $\qe$, these
quantities obey
\bea
\label{eqstateMR}
	\frac{\qo_s zt}{M^\qd} &=& (t_c/t)^{1/\qe}\left(1
	-2\qe\nu
	\frac{1-t/t_c}{M^{1/\qb_0}}\right)^{1/\qe} \\
	\frac{\qo_s zt}{R^\qk} &=& (t_c/t)^{1/\qe}
	\left(1-\frac{1-t/t_c}{R^{1/2\qe\nu}}\right)^{1/\qe}\nn
\eea
with the following relations between the exponents
\bea
	D\nu=\qb_0(\qd+1) \hskip0.5cm &;& \hskip0.5cm \qk=\qb_0\qd/(2\qe\nu).
\eea
The equation of state for the conductivity is obtained by replacing
$R\!\naar\! R_\qs$ and $\qo_s\!\naar\! i\qo$ in (\ref{eqstateMR}).

\ns{Specific heat}
\label{secspecheat}

In order to make contact with an important quantity like the specific
heat, we will first evaluate the derivative $\prt{\cal F}/\prt\ln T$
from the fermionic path integral for the case of free particles, and
compare the results to those obtained from $Q$-field theory. By the
rules of statistical mechanics we have
\be
	-\frac{\prt{\ln Z}}{\prt\ln\qb}=N_r \qb( \bar E-\mu \bar N)
\label{specheat}
\ee
where $\bar E$ is the average energy and $\bar N$ the average number of
electrons. Since the specific heat is defined as the change of 
$\bar E$ with varying $T$ for a fixed number of particles, we still
need some procedure that eliminates the explicit $\mu$-dependence in
(\ref{specheat})

\subsection{Free particles}
\label{secfreepart}

From the original fermionic path integral we obtain directly
\be
       V_D \fr{\prt{\cal F}}{\prt\ln\qb}=\sum_{n}i\qo_{n}\;\Tr
        \fr{1}{i\qo_{n}+\mu-H_{0}}
\label{freqsum}
\ee
with $\ln Z\!\!=\!\! -V_D N_r{\cal F}$. 
The `Tr' stands for the trace over position space and $V_D$ is
the volume of the system.
By writing the frequency sum as a contour integral in the usual manner,
\be
        V_D \fr{\prt{\cal F}}{\prt\ln\qb}=-\fr{\qb}{2\pi i}\oint\!\!
        dz\;\fr{z}{e^{\qb z}+1}\;\Tr\fr{1}{z+\mu-H_{0}},
\ee
we obtain the more transparent expression
\be
        \qb V_D
        \int_{-\infty}^{\infty}\!\! d\epsilon\; 
        \frac{\epsilon}{e^{\qb\epsilon}+1}\qr(\epsilon)
\label{rho(eps)}
\ee
where $\qr(\epsilon)$ is the density of states at energy 
$\mu\!+\!\epsilon$ and $V_D$
the volume of the system.
The equivalence between Eq (\ref{specheat}) and (\ref{rho(eps)}) is now easily
established. By splitting Eq (\ref{rho(eps)}) into zero-$T$ and finite-$T$
parts we finally obtain
\bea
        \fr{\prt{\cal F}}{\prt\ln\qb} & = & \qb(f_{0}+f_{T})
        \nn \\
        & = &  \qb \int_{-\infty}^{0}\!\! d\epsilon\; \epsilon\qr(\epsilon)
        + \qb \int_{0}^{\infty}\!\! d\epsilon\; \frac{\epsilon}{e^{\qb\epsilon}+1}
        [\qr(\epsilon)+\qr(-\epsilon)] \nn \\
        & = &
        \qb\left(-\int_{-\infty}^{\mu}\!\! dE\; n(E)
        +\int_{0}^{\infty}\!\! d\epsilon\;
        \frac{\epsilon}{e^{\qb\epsilon}+1} 
        [\qr(\epsilon)+\qr(-\epsilon)] \right)
\label{split}
\eea
with $n(E)$ the total number of states with energy less than $E$ 
($\prt n(E)/ \prt E$ is equal to the density of states).
The $f_{T}$ determines the well known specific heat of the free
electron gas for low temperatures
($c_{v}\! =\! \prt f_T/\prt T \! =\!\qg T$). 
Next, we wish to redo the various steps
which take us from Eq (\ref{freqsum}) to (\ref{split}), but now
starting
from the effective theory in $Q$. We obtain, instead of 
Eq (\ref{freqsum})
\be
        \fr{\prt{\cal F}}{\prt\ln\qb}=\pi\qr_0 N_r\inv\;
        \tr\expec{\qo Q} = 2\pi\qr_0
        \sum_{n\geq 0}\;\!\!\!^{'}\qo_{n}
\label{sumfromQ}
\ee
where $\qo_n\!=\!\pi(2n+1)/\qb$ 
and the prime on the summation indicates that the sum involves a
cutoff. By taking a simple exponential form, i.e. replacing 
Eq~(\ref{sumfromQ}) by
\be
        2\pi \qr_0\sum_{n\geq 0}
        \qo_{n}e^{-\qo_{n}\qt_{0}},
\label{cutoff}
\ee
and by redoing the various steps which take us from Eq~(\ref{freqsum})
to (\ref{split}), one is easily convinced that the low $T$ behaviour
of the specific heat is identically the same as was obtained before.
To be more specific, Eq.~(\ref{cutoff}) gives
\be
	f_0=\int_0^\infty\! d\varepsilon \; \varepsilon\qr_0 
	e^{-\varepsilon\qt_0}
	\hskip5mm {\rm and} \hskip5mm
	f_T=\int_0^\infty\!
	d\varepsilon\frac{\varepsilon}{e^{\qb\varepsilon}+1}2\qr_0. 
\ee
(All the dependence on the cutoff $\qt_{0}$ is absorbed in the
zero-$T$ part of Eq~(\ref{split}), i.e. $f_{0}$, and this quantity 
is of secondary interest). Hence, the low-$T$ specific heat of the
electron gas is correctly retained by the effective $Q$-field formalism.

\subsection{Coulomb interactions; quasiparticles}
\label{secCoulquasi}

Next we embark on the Coulomb interaction problem. 
We obtain from the Finkelstein action, instead of
Eq~(\ref{sumfromQ}),
\be
        \fr{\prt{\cal F}}{\prt\ln\qb}=\fr{\pi}{2\qb}z_0\left< {\sum_{n}}'
        \tr [\Ian,Q][\Iamn,Q]
        \right>.
\label{Finkel}
\ee
Besides the renormalizable quantity in the $Q$-matrix fields, there
are also other, $Q$-independent, contributions to (\ref{Finkel}) that
are still left over from the underlying theory of longitudinal modes
(see [I]). These contributions, however, are of a Fermi liquid type
and, hence, of secondary interest. Using the results of
Section~\ref{secscaling}, we write (\ref{Finkel}) as follows
\be
        -2 z_{0}\sum_{n>0}\qo_{n}
        M(\qo_{n})e^{-\qo_{n}\qt_{0}}
\label{FinkelM}
\ee
where  the sum is now over bosonic Matsubara frequencies
$\qo_n\!=\! 2\pi n/\qb$ and, as before, we introduced an
arbitrary $\qt_{0}$ for convergence 
purposes. Notice that the quantity $M(\qo_{n})$,
in contrast to the density of states $\qr_0$ in the expression for
the free electron problem (\ref{cutoff}), now acquires nontrivial
$\qo_{n}$-dependence.
We proceed by evaluating the discrete sum in (\ref{FinkelM})
as a contour integral. Repeating the same steps which lead to
(\ref{split}) we obtain
\be
        \fr{\prt{\cal F}}{\prt\ln\qb} = 
        \qb(f_{0}+f_{T}),
\ee
where
\be
        f_{0}  =  -\fr{z_0}{\pi}
	\int_0^\infty\!\! d\epsilon\; \epsilon M(\epsilon)
        e^{-\epsilon\qt_{0}}
\ee
and $f_T$ can be written as
\be
	f_T=\fr{z_{0}}{\pi}\int_{0}^{\infty}\!\! d\epsilon\;
        \frac{\epsilon}{e^{\qb\epsilon}-1}\qr_{qp}(\epsilon)
\label{omegaT}
\ee
with
\be
	\qr_{qp}(\epsilon)=e^{i\epsilon \qt_0}
	M(-i\epsilon)+e^{-i\epsilon \qt_0}M(i\epsilon).
\ee
As before (Eq \ref{split}), the specific heat is obtained as
$c_{V}\! =\! \prt f_{T}/ \prt T$. The quantity $\qr_{qp}(\epsilon)$ is
now identified 
as the quasiparticle density of states of the interacting system.

The expression for $f_{T}$, Eq~(\ref{omegaT}), is the most important
result of this section and we make use of the results for 
$M(\qo_{n})$, section \ref{secscaling}, 
in order to extract the behavior of the specific heat.
First, in the metallic phase in $2\! +\! 2\qe$ dimensions, the arbitrary
cutoff factor containing $\qt_{0}$ in $f_{T}$ does not contribute
to the leading behaviour of $c_{V}$ at low $T$, and $\qt_{0}$ can
be safely put to zero. This can be seen most simply from the behaviour
near the fixed points at $t\! =\! 0$,

\bea
        t\rightarrow 0  \hskip1cm & (\mbox{Fermi liquid phase})& \nn \\
        M(\qo_{n})\approx 1 \hskip1cm 
	& 
	\qr_{qp}\rightarrow 2\cos(\epsilon\qt_{0})
        \approx 2 & 
	\hskip1cm 
	c_{v}\rightarrow \qg_{0}T \nn
\eea
and at $t\! =\! t_{c}$, in which
case we have 
\bea
\label{rhoqp}
        t= t_{c} \hskip1cm
        & (\mbox{critical phase})\hskip1.5cm & \\
        M(\qo_{n})\approx |\qo_{n}|^{1/ \qd}  \hskip1cm & 
        \qr_{qp}\approx |\epsilon|^{1/
        \qd}2\cos(\fr{\pi}{2\qd}+\epsilon\qt_{0}) 
        & \hskip1cm 
	c_{v}\rightarrow \qg\st T^{1+1/ \qd}. \nn
\eea
Hence, approaching the metal-insulator transition from the metallic side, the
(bosonic) quasiparticle density of states develops a (Coulomb) gap
and the exponent $1/ \qd$ is of order 
$\qe\! =\!(D\! -\! 2)/2$.
More generally, we have the following scaling result,
\be
        \qr_{qp}(\epsilon)=M_{0}G(\epsilon z\xi^{D}M_{0})=M(-i\epsilon)+M(i\epsilon)
\label{scaling}
\ee
which may be obtained in closed form following the results for $M$ of
section~\ref{secscaling}.

\begin{figure}
\begin{center}
\setlength{\unitlength}{1mm}
\begin{picture}(90,50)(0,0)
\put(0,0)
{\epsfxsize=90mm{\epsffile{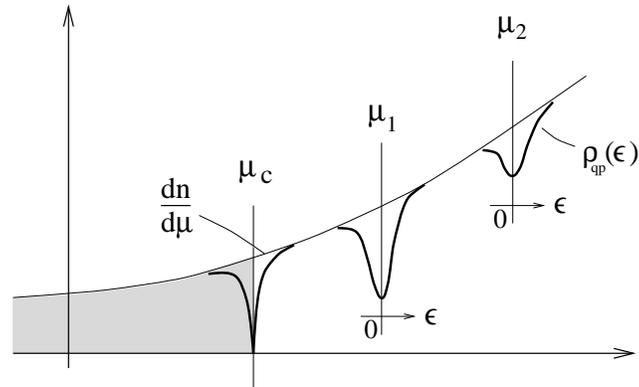}}}
\end{picture}
\caption{The quasiparticle density of states in $2\!+\!2\qe$
dimensions for different values of the chemical potential ($\mu_1$,
$\mu_2$ and $\mu_c$). The critical value $\mu_c$ separates the
insulating phase (shaded region) from the metallic phase (unshaded
region); see text.}
\label{figqpDOS}
\end{center}
\end{figure}

For completeness we give a sketch of $\qr_{qp}(\epsilon)$ as compared to
the thermodynamic density of states $\prt n/\prt\mu$, 
Fig.~\ref{figqpDOS}.
The $\prt n/\prt\mu$ does not enter the effective $Q$-theory
and is taken as a regular function of the chemical potential, since it
is solely determined by the underlying theory with the $P$ matrix
field variables.
Figure~\ref{figqpDOS} indicates that $\qr_{qp}(\epsilon)$ as a whole moves along as
one varies the chemical potential $\mu$ with $\epsilon$ really standing for
the thermal energy fluctuations for a given, fixed $\mu$. The
effective band width ($\qD\epsilon$) for quasiparticle excitation is on the
order of $\hbar/\qt_0$, i.e. the characteristic energy scale below
which the Coulomb effects become noticeable. It is, of course, understood
that for free electrons there is no distinction between $\qr(\epsilon)$
(\ref{split}) and $\prt n/\prt\mu$, and the quasiparticles are the
electrons themselves.

The formation of a gap in $\qr_{\rm qp}$ is reminiscent of the
Efros-Shklovskii Coulomb gap\cite{EfrosShklovskii} in the tunneling
density  of states in the localized phase. 
However, since the tunneling density of states is
not a gauge invariant object\cite{b8}, the relation between the two is
not obvious. 
In order to make contact with the heuristic approach by Efros and
Shklovskii\cite{EfrosShklovskii}, it 
is necessary to encompass the limitations of the perturbative
renormalization group and find a way to penetrate into the regime of strong
coupling. We elaborate further on this point in Section~D.

\subsection{Crossing over between free electrons and quasiparticles}
\label{seccross}

Let us next come back to our earlier discussion 
(section \ref{secbackmethod}), where
we replaced the original problem with Coulomb interactions (\ref{SU}) by
a simpler one (\ref{Sback}) introducing the parameter $c_0$ (or $\qa$). We
argued that since the Coulomb term $S_{\rm U}$ is irrelevant we obtain
the same results by putting 
$c_0\! =\! 1$ ($\qa\! =\! 0$) in the simpler theory.
We are now in a position to appreciate the fact that the two theories
really stand for entirely different physical scenarios by which the
free electron and Coulomb theories are related.

As already mentioned before, the parameter $\qa$ in the simple theory
can (loosely) be interpreted in terms of the `range' of the
electron-electron interactions, and the results indicate that infinite
range interactions 
($\qa\! =\! 0$) and finite range interactions ($\qa\! >\!0$)
belong to different universality classes. On the other hand, the
Coulomb term $S_{\rm U}$ in the original theory really interpolates
between the action for free particles at high momenta and the
Finkelstein theory which only appears in the limit of large distances
(relative to the screening length).

This means that the true, physical theory predicts {\em free electron}
behaviour with Fermi-Dirac statistics at {\em high} $T$ (\ref{split})
and {\em quasiparticle} behaviour with scaling and Bose-Einstein
statistics at {\em low} $T$ (\ref{rhoqp}). This crossover mechanism
for which the Coulomb term $S_{\rm U}$ is responsible obviously
applies to other physical quantities such as $\qs_{xx}$.
This mechanism is extremely important, since it has been
experimentally shown to apply to the more complicated metallic phases
of the quantum Hall regime as well. More specifically, the transport
data taken from low mobility heterostructures follow the well known
free electron behaviour at high $T$ 
(4K$\leq T  < $20K),
\be
	\qs_{ij}(T)=\int\! dE\; \fr{\prt f(T)}{\prt E}\qs_{ij}^0(E)
\ee
Here $f(T)$ stands for the Fermi-Dirac distribution
(see \ref{freqsum}) and $\qs_{ij}^0$ are the mean field conductances
with varying energy $E$.

This behaviour crosses over smoothly but rapidly into a scaling
behaviour in $T$, which was observed at lower temperatures
(20mK$\leq T<$4K) only. This, then, indicates that the
complete theory as given by (\ref{SQA}) has all the ingredients necessary
to describe the complex phenomenon of plateau transitions in the
quantum Hall regime.

\subsection{Strong coupling aspects}
\label{secstrongcoupl}

The analogy with the Heisenberg ferromagnet as discussed in previous
sections naturally supposes that one can take the theory one step
further by extending the ferromagnetic language to include the
insulating phase as well. For example, for ordinary ferromagnets we
know that, since there is no phase transition in the symmetric phase,
the magnetization $M(\qo_n)$ should become a regular and odd function
of the `external field' $\qo_n$. 
From the renormalization group it can be shown that such a condition on
$M(\qo_n)$ implies that the $R(\qo_n)$ (but not $R_\qs$!) is regular as
well but it must be an even function of $\qo_n$.
This analyticity statement (Griffith
analyticity) would imply in our case, however, that the quasiparticle
density of 
states collapses, i.e. (\ref{scaling}) vanishes to all orders in $\epsilon$!

This oddly looking conclusion actually has an extremely important
physical significance and it can easily be understood on the basis of
our discussion of ${\cal F}$-invariance. First we remark that a vanishing
$\qr_{qp}(\epsilon)$ is a direct consequence of working with an action where
the Coulomb part $S_{\rm U}$ has been replaced by a symmetry braking
term $S_\qa$ and where the `external' symmetry breaking $\qa$ has been
put equal to zero. Notice that for Fermi level quantities like
$\qr_{qp}(\epsilon)$ and $\qs(\qo)$ the role of $\qa$ itself is analogous to
the role of the magnetic field in the Heisenberg ferromagnet.

This is evident from the linear response procedure, where
the perturbing vector potential can be considered as a `generator' of
${\cal F}$-symmetry which is broken by the external $\qa$-field. Hence we
should expect that the quantities $\qr_{qp}$ and $\qs$, which are a
direct measure of the low energy excitations of the electron gas, become
regular in $\qa$ as one enters deeply into the insulating phase. In
other words, they become zero by putting $\qa$ equal to zero.

The important conclusion that can be drawn from this discussion is that
the Coulomb term $S_{\rm U}$ (\ref{SU}) can no longer be regarded as
`irrelevant' but, instead, it is going to completely determine the
physics of the insulating phase.

We conclude this section by giving a brief digression on the general
significance of the Coulomb part of the action. This, then, serves as
a starting point for a more extended renormalization group analysis
which will be reported elsewhere. 
For this purpose, 
let us go back to the original theory (\ref{SQA}) and compute the
response to an external field $A_\mu$. At a tree level, the result is
given by (see \cite{3A})
\be
	S[A]=-\qs_{xx}^0\sum_{\qa,n>0}\int\!\frac{d^2 q}{(2\pi)^2}
	n\;\;\overline{(z_i)^\qa_n(q)}\left[\qd_{ij}
	-\frac{q_i q_j}{q^2+\qk^2 nU\inv(q)}
	\right](z_j)^\qa_n(q)
\label{responsexx}
\ee
where $\vec z^\qa_n$ is defined as 
$\vec A^\qa_n \! -\! i\nabla(A_\qt)^\qa_n/\qo_n$.
The result (\ref{responsexx}) can be used to show that the nature of
particle transport in metals changes depending on what length scale
one probes the system at. From the definition for the particle density $n$
\be
	-\qb n^\qa_m(q)=\frac{\qd S[A_\mu]}{\qd (A_\qt)^\qa_{-m}(-q)}
\label{density}
\ee
we obtain, following (\ref{responsexx})
\be
	\left[\qo_m+\fr{1}{4}\qs_{xx}^0 q^2 U(q)\right]n^\qa_m(q)
	=i\vec q\cdot (\vec\jmath_{\rm ext})^\qa_m(q).
\label{deltaFsp}
\ee
The $\vec\jmath_{\rm ext}$ is the current density induced by external
fields. 
We have obtained this current density by using
$\vec z^\qa_m\! =\!i\vec E^\qa_m/\qo_m$ and 
$\vec\jmath_{\rm ext}\! =\!\fr{\qs_{xx}^0}{2\pi}\vec E$.
For large momenta we have 
\be
	U(q)\approx \fr{2}{\pi}\qr\inv \hskip5mm
	{\rm as} \hskip5mm
	|q|\naar\infty
\ee
and (\ref{deltaFsp}) becomes
the expression for particle conservation in a diffusive system driven
by an external potential,
\be
	\prt_t n_c+\nabla\cdot (\vec\jmath_{\rm ext}
	+\vec\jmath_{\rm diff})=0
\label{diffcontinuity}
\ee
where 
$n_c\!=\!-n$ is the charge density,
$\vec\jmath_{\rm diff}\! =\! -D\nabla n_c$ with 
$D\! =\!\qs_{xx}^0/(2\pi\qr)$ the diffusion constant.
On the other hand, in the small $q$ limit one has
\be
	U(q)\naar \fr{2}{\pi} U_0(q) \hskip5mm
	\mbox{as} \hskip5mm
	|q|\naar 0.
\ee
We now write 
$\vec E_c(q)\! =\! - i\vec q U_0(q) n_c(q)$ or, in position space,
\be
	\vec E_c(x)=-\nabla\intd{^2 x'}U_0(x,x')n_c(x')
\ee
for the local electric field as a result of the interaction with all
the other electrons in the system.
We  now have, instead of
(\ref{diffcontinuity}), 
\be
	\prt_t n_c+\nabla\cdot(\vec\jmath_{\rm ext}+\vec\jmath_c)=0
\label{Coulombcontinuity}
\ee
where $\vec\jmath_c\! =\!\fr{\qs_{xx}^0}{2\pi} \vec E_c$.

Equation (\ref{Coulombcontinuity}) indicates that `detailed balance'
no longer stands for mutually compensating electric and `diffusive'
currents. Rather, the external field is canceled by the internally
generated electric field due to the Coulomb forces.
The length scale at which one crosses over from the diffusion
dominated regime (\ref{diffcontinuity}) to the `Coulomb driven' regime
(\ref{Coulombcontinuity}) is given by the Debye static screening
length $(2\pi\qr)\inv$.

Let us next come back to the subject of linear response. It is clear
that for the computation of conductivities one generally has to take the
limit $q\!\naar\! 0$ first and then $\qo\!\naar\! 0$. Notice that the
abovementioned complications with with the limit 
$\qa\!\naar\! 0$ are, in
fact, foreshadowed by the response at tree level (\ref{responsexx}). For
instance, working in a theory where $U(q)$ is replaced by $\qa\inv$ (as
was done for renormalization group purposes) then linear response leads
to the correct result provided $\qa\!\naar\! 0$ in the end. As one
approaches the insulating phase, however, it seems from (\ref{responsexx})
that the renormalization group result 
($\qs\!\propto\! \qa\xi^2\qo$)
is indeed the only way of obtaining a smooth $\qa\!\naar\! 0$
limit. 
This indicates that the theory with $\qa\!=\!0$ develops an energy
gap. However, in the presence of $U(q)$ it is more likely that
$\qs\!\propto\!\xi\qo$ in the insulating phase, which means that the
dominant energy scale is now determined by the Coulomb potential.


\ns{Conclusion}
\label{secconclusion}

In this paper we have reported the results of a perturbative
renormalization group analysis (to two loop order) of the Finkelstein
theory of localization and interaction effects. 
This theory has fundamental significance for the quantum Hall effect
and it is part of a unifying action as proposed in our previous work
\cite{3A}.

We have shown that the infrared behaviour of the theory 
(i.e. the limit $h_0\!\naar\! 0$)
can only be
extracted from a limited class of 
(${\cal F}$-invariant) correlation functions. This insight
enabled us to identify several new quantities
in the theory, such as the
`order parameter' of the metal-insulator transition in $2\! +\! 2\qe$
dimensions and the `Coulomb gap' which enters through the (bosonic)
quasiparticle density of states
into the expression for the specific heat.

${\cal F}$-invariance has also fundamental consequences 
for the insulating phase, where ordinary perturbation theory is no
longer valid. To this end, we exploited the analogy of the
Finkelstein theory with the more familiar
theory of the classical Heisenberg ferromagnet. We have
shown that the transport problem in this case must be completely
dominated by the Coulomb part of the action ($S_{\rm U}$) which is
usually discarded on the basis of naive scaling dimensions.
The appropriate way of demonstrating this is by computing the
anomalous dimension of the operators in $S_{\rm U}$. This requires a 
more extended renormalization group program than the one presented in
this paper.
Progress in this direction, along with the possible consequences for
the plateau transition in the quantum Hall regime, will
be reported elsewhere.

\vskip1cm

\noindent
{\bf ACKNOWLEDGEMENTS}\newline
This research was supported in part by
INTAS (grant \#96-0580).

\vskip1cm
\scez\renewcommand{\theequation}{A\arabic{equation}}
\noindent
{\Large\bf Appendix}

\vskip0.4cm
\noindent

In this appendix we present the final expressions for the various
terms in (\ref {2.6.1})-(\ref{2.6.8}) together with some calculational details.
The calculation of (\ref{2.6.1})-(\ref{2.6.5}) is straightforward
because the
internal momenta decouple, and we just present the results (only pole
terms in $\qe$), taking the limit 
$\alpha\!\naar\! 0$ wherever possible:
\begin{eqnarray*}
	\int_{p,q}D_{q}(s)D_{p}(s) &\rightarrow
	& \qO_D^2 z_0 t_0^2 \frac{h_0^{4\qe}}{\qe^2}(1+ 
	\widehat{\omega }_{s})^{2\varepsilon }
	\left[ \frac{\pi \varepsilon }{\sin \pi
	\varepsilon }\right] ^{2} \\
	\qk^2 z_0\sum_{l>s}\int_{p,q}D_{q}(s)D_{q}^{c}(s)D_{p}(s+l)
	&\rightarrow &\Omega _D^{2}h_{0}^{4\varepsilon }\cdot 
	\frac{1}{\varepsilon^{2}(1+\varepsilon )}
	\left\{ (1+2\widehat{\omega }_{s})^{1+\varepsilon }\frac{(1+ 
	\widehat{\omega }_{s})^{\varepsilon }-1}{\widehat{\omega}_{s}}
	\right\} \\
	\qk^2 z_0\sum_{l>s}\int_{p,q}D_{q}(s+l)D_{p}(l)D_{p}^{c}(l)
	&\rightarrow &\Omega _D^{2}h_{0}^{4\varepsilon }\cdot 
	\left\{ \frac{{\rm I}_{3}(\varepsilon )}
	{2\varepsilon ^{2}}\frac{\alpha ^{-\varepsilon }-1}{\varepsilon }+ 
	\frac{1}{\varepsilon }\int_{0}^{1}\frac{dx}{x}
	\ln(1+x\widehat{\omega }_{s})\right\} \\
	\qk^2 z_0\sum_{l>s}\int_{p,q}D_{q}^{c}(s)D_{p}(l)D_{p}^{c}(l) 
	&\rightarrow &\Omega _D^{2}h_{0}^{4\varepsilon }\cdot 
	\left\{ -\frac{{\rm I}_{3}(\varepsilon )}{\varepsilon ^{2}}
	\ln \alpha \cdot 
	\frac{\left[\Gamma (1-\varepsilon )\right]^{2}}
	{\Gamma (1-2\varepsilon )}\right.
	\\ &&
	\left. +\frac{1}{\varepsilon } 
	\int_{0}^{1}\frac{dx}{x}\ln (1+x\widehat{\omega }_{s})\right\} \\
	\qk^2 z_0\sum_{l>s}\int_{p,q}D_{q}(s)D_{q}^{c}(s)D_{p}^{c}(l) 
	&\rightarrow &\Omega _D^{2}h_{0}^{4\varepsilon }\cdot 
	\left\{ \frac{1}{\alpha \varepsilon ^{2}}
	\frac{(1+z)^{\varepsilon }-(1+\alpha z)^{\varepsilon }}
	{z(1-\alpha )(1+\varepsilon )}\right. \\
	&&\left.
	+\frac{1}{\varepsilon ^{2}}
	\left[ (1+\widehat{\omega }_{s})^{\varepsilon }-1\right] \right\} ,
\end{eqnarray*}
where we have defined 
$\widehat{\qo}_s\! =\! h_s^2/h_0^2$
and ${\rm I}_{3}(\varepsilon )$ stands for the combination
\[
	{\rm I}_{3}(\varepsilon )=
	\frac{\Gamma (1-2\varepsilon )}{\left[ \Gamma
	(1-\varepsilon )\right] ^{2}}
	\left[ \frac{\pi \varepsilon }{\sin \pi \varepsilon}\right]^{2}
	\approx 1+\frac{\pi ^{2}}{2}\varepsilon^{2}
	+{\cal O}(\varepsilon ^{3}).
\]
The calculation of (\ref{2.6.6}) is more difficult but standard. As
far as we are interested only in poles in $\varepsilon$ we can replace
the sum over $l$ by
an integral and put the lower limit to zero rather that $\omega _{s}$:
\[
	\fr{2\pi}{\qb}\sum_{l>s}\rightarrow 
	\int_{0}^{\infty }d\omega _{l}+{\cal O}(\varepsilon^{0}).
\]
Using the Feynman trick one can write
\bea
	&& \qk^2 z_0 
	\sum_{l>s}\int_{p,q}D_{q}^{c}(s)D_{p}^{c}(l)D_{{\bf p+q} 
	}(s+l) = 4\fr{z_0}{\qs_0}\int_{0}^{\infty }d\omega_{l}
	\int_{p,q}\int_{0}^{1}\prod_{i=1}^{3}dx_{i} \times\nn\\
	&& \times
	\frac{2\cdot \delta
	(1-\sum_{i}x_{i})}{\left[ h_{0}^{2}+q^{2}x_{12}+p^{2}x_{23}+2{\bf pq} 
	x_{3}+\qk^2 s(\alpha x_{1}+x_{3})+\qk^2 l(\alpha
	x_{2}+x_{3})\right]^{3}},  
\label{A1}
\eea
where 
$x_{13}\! =\! x_{1}\! +\! x_{2}$ and 
$x_{23}\! =\! x_{2}\! +\! x_{3}$. By shifting   
$p\!\naar\! p\! -\! q x_{3}/x_{23}$ we can decouple $p$ and $q$ in the
denominator, after which we are able to perform the integrals over
$p,q$ and $\omega_{l}$, with the result
\begin{equation}
	-\frac{{\rm I}_{3}(\varepsilon )}{2\varepsilon }
	\Omega _D^{2}h_{0}^{4\varepsilon}
	\int_{0}^{1}\prod_{i=1}^{3}dx_{i}\frac{\delta(1-\sum_{i}x_{i})}
	{\left[x_{1}x_{2}+x_{1}x_{3}+x_{2}x_{3}\right]^{1+\varepsilon}}
	\cdot \frac{\left[ 1+\widehat{\omega }_{s}
	(\alpha x_{1}+x_{3})\right] ^{2\varepsilon }}{(\alpha x_{2}+x_{3})}.  
\label{A2}
\end{equation}
Introducing new variables $x\in \left[ 0,\infty \right] $ and $u\in \left[
0,1\right] $ as follows
\[
	x_{1}=\frac{x}{x+1} \hskip0.3cm ; \hskip0.3cm 
	x_{2}=\frac{u}{x+1} \hskip0.3cm ; \hskip0.3cm 
	x_{3}=\frac{1-u}{x+1},
\]
formula (\ref{A2}) can be written in the form
\begin{equation}
	-\frac{{\rm I}_{3}(\varepsilon )}{2\varepsilon }
	\Omega _D^{2}h_{0}^{4\varepsilon}\int_{0}^{1}
	\frac{du}{\alpha u+1-u}\int_{0}^{\infty }dx\frac{\left[ 1+ 
	\widehat{\omega }_{s}(1-u)+x(1+\alpha \widehat{\omega }_{s})\right]
	^{2\varepsilon }}{\left[ x+u(1-u)\right] ^{1+\varepsilon }}.  
\label{A3}
\end{equation}
Using a well-known integral representation for the hypergeometric
function, the $x$-integral can be performed and we get the following
expression for (\ref{A1})
\begin{equation}
	-\frac{{\rm I}_{3}(\varepsilon )}{2\varepsilon }
	\Omega _D^{2}h_{0}^{4\varepsilon}\left[ {\rm I}_{1}
	-\frac{\left[ \Gamma (1-\varepsilon )\right] ^{2}}{\Gamma
	(1-2\varepsilon )}\cdot {\rm I}_{2}\right],
\label{A4}
\end{equation}
where 
\begin{eqnarray}
	{\rm I}_{1} &=&
	\int_{0}^{1}du\frac{\left[ u(1-u)\right] ^{-\varepsilon }}
	{\alpha u+1-u}\left[ 1+\widehat{\omega }_{s}(1-u)\right] 
	^{2\varepsilon }\ _{2} 
	{\rm F}_{1}(1,-2\varepsilon ;1-\varepsilon ;
	\frac{u(1-u)(1+\alpha \widehat{\omega }_{s})}
	{1+\widehat{\omega }_{s}(1-u)})  
\label{A5} 
	\\
	{\rm I}_{2} &=&2\int_{0}^{1}\frac{du}{\alpha u+1-u}
	\left\{ 1+(1-u)\left[\widehat{\omega }_{s}
	-u(1+\alpha \widehat{\omega }_{s})\right] \vphantom{H^H}
	\right\}^{\varepsilon}.
\label{A6}
\end{eqnarray}
We are interested in the limit $\qa\!\naar\! 0$, which corresponds to
the case of real electrons. But if one puts 
$\alpha\! =\! 0$ in the above
formulae then (\ref{A6}) becomes divergent on the upper limit. 
The integral (\ref{A5}) will be proportional to $\varepsilon ^{-1}$
making the whole 
contribution (\ref{A1}) proportional to $\varepsilon ^{-3}$ and causing
the renormalization group
equation to become singular.
For this reason we first rewrite (\ref{A5}) in the
following form
\begin{eqnarray}
{\rm I}_{1} &=&\int_{0}^{1}du\frac{\left[ u(1-u)\right] ^{-\varepsilon }}{ 
\alpha u+1-u}  \label{A7.1} \\
&&+\int_{0}^{1}du\frac{\left[ u(1-u)\right] ^{-\varepsilon }}{\alpha u+1-u} 
\left\{ \left[ 1+\widehat{\omega }_{s}(1-u)\right] ^{2\varepsilon }-1\right\} 
\label{A7.2} \\
&&+\int_{0}^{1}du\frac{\left[ u(1-u)\right] ^{-\varepsilon }}{\alpha u+1-u} 
\left[ 1+\widehat{\omega }_{s}(1-u)\right] ^{2\varepsilon }
\times\nn\\
&&\hskip1cm\times
\left\{  _{2}{\rm F} 
_{1}(1,-2\varepsilon ;1-\varepsilon ;\frac{u(1-u)(1+\alpha \widehat{\omega }_{s})}{ 
1+\widehat{\omega }_{s}(1-u)})-1\right\}.
\label{A7.3}
\end{eqnarray}
The first integral (\ref{A7.1}) can be done exactly while in the other two
we can safely put 
$\alpha\! =\! 0$, because their integrands behave smoothly when 
$u\!\naar\! 1$ even for 
$\alpha\! =\! 0$. Thus ${\rm I}_{1}$ can be written in
the form
\begin{eqnarray*}
{\rm I}_{1} &=&\frac{1}{\varepsilon }\left\{ \alpha ^{-\varepsilon }\frac{\pi
\varepsilon }{\sin \pi \varepsilon }-\frac{\left[ \Gamma (1-\varepsilon )\right] ^{2} 
}{\Gamma (1-2\varepsilon )}\right\} +{\cal O}(\alpha ^{1-\varepsilon }) \\
&&+\int_{0}^{1}du\frac{\left[ u(1-u)\right] ^{-\varepsilon }}{1-u}\left\{
\left[ 1+\widehat{\omega }_{s}(1-u)\right] ^{2\varepsilon }-1\right\}  \\
&&+\int_{0}^{1}du\frac{\left[ u(1-u)\right] ^{-\varepsilon }}{1-u}\left[ 1+ 
\widehat{\omega }_{s}(1-u)\right] ^{2\varepsilon }\ \left\{ _{2}{\rm F} 
_{1}(1,-2\varepsilon ;1-\varepsilon ;\frac{u(1-u)}{1+\widehat{\omega }_{s}(1-u)} 
)-1\right\}. 
\end{eqnarray*}
Making an $\varepsilon$-expansion and then using 
the following asymptotic formula for the hypergeometric $_2{\rm F}_1$
function, which follows directly from its definition,
\[
_{2}{\rm F}_{1}(1,-2\varepsilon ;1-\varepsilon ;z)=1+2\varepsilon \ln (1-z)+{\cal O} 
(\varepsilon ^{2})
\]
we get the final answer for ${\rm I}_{1}$ 
\begin{eqnarray}
{\rm I}_{1} &=&\frac{1}{\varepsilon }\left\{ \alpha ^{-\varepsilon }\frac{\pi
\varepsilon }{\sin \pi \varepsilon }-\frac{\left[ \Gamma (1-\varepsilon )\right] ^{2} 
}{\Gamma (1-2\varepsilon )}\right\}   \label{A8} \\
&&+2\varepsilon \int_{0}^{1}\frac{du}{1-u}\ln \left[ 1+(1-u)(\widehat{\omega } 
_{s}-u)\right] +{\cal O}(\varepsilon ^{2}).  \nn
\end{eqnarray}
The integral ${\rm I}_{2}$ (\ref{A6}) can be treated in the same way
\begin{eqnarray}
{\rm I}_{2} &=&2\int_{0}^{1}\frac{du}{\alpha u+1-u}+2\int_{0}^{1}\frac{du}{ 
1-u}\left\{ \left[ 1+(1-u)\left( \widehat{\omega }_{s}-u\right) \right]
^{\varepsilon }-1\right\}   \label{A9} \\
&=&-2\ln \alpha +{\cal O}(\alpha \ln \alpha )  \nn \\
&&+2\varepsilon \int_{0}^{1}\frac{du}{1-u}\ln \left[ 1+(1-u)(\widehat{\omega } 
_{s}-u)\right] +{\cal O}(\varepsilon ^{2}).  \nn
\end{eqnarray}
Substitution of (\ref{A8}) and (\ref{A9}) into (\ref{A3}) gives the final
answer for the pole terms in (\ref{A1})
\bea
&& \qk^2 z_0 \sum_{l>s}\int_{p,q}D_{q}^{c}(s)D_{p}^{c}(l)D_{{\bf p+q} 
}(s+l) \nn\\
&&
=\Omega _D^{2}h_{0}^{4\varepsilon }\cdot \frac{{\rm I}_{3}(\varepsilon )}{ 
2\varepsilon }\left\{ -\frac{2\ln \alpha }{\varepsilon }\frac{\left[ \Gamma
(1-\varepsilon )\right] ^{2}}{\Gamma (1-2\varepsilon )}-\frac{1}{\varepsilon }\frac{ 
\alpha ^{-\varepsilon }-1}{\varepsilon }\cdot \frac{\pi \varepsilon }{\sin \pi
\varepsilon }+\frac{\pi ^{2}}{3}\right\} .
\eea
The calculation of (\ref{2.6.7}) follows the same line but is more lengthy. We
will not present it here but give only some comments. In this case the
second power of the combination 
$\alpha u\! +\! 1\! -\!u$ appears in the denominator at
the stage (\ref{A4}). It is useful then to perform integration by parts,
reducing the power by one. The rest is similar to the calculations of (\ref
{A1}). The final answer is
\begin{eqnarray*}
&& -\qk^4 z_0 c_0\sum_{l>s}l\cdot\int_{p,q}D_{p}(s)D_{p 
}^{c}(s)D_{q}^{c}(l)D_{{\bf p+q}}(s+l) \\
&& \longrightarrow
\Omega
_D^{2}h_{0}^{4\varepsilon }\cdot \frac{{\rm I}_{3}(\varepsilon )}{2\varepsilon ^{2}} 
\left\{ -\frac{2}{\alpha }\cdot \frac{\left[ \Gamma (1-\varepsilon )\right] ^{2} 
}{\Gamma (1-2\varepsilon )}\cdot \frac{(1+z)^{\varepsilon }-(1+\alpha z)^{\varepsilon
}}{z(1-\alpha )(1+\varepsilon )}\right.  \\
&&-\left. \frac{\alpha ^{-\varepsilon }-1}{\varepsilon }\cdot \frac{\pi \varepsilon }{ 
\sin \pi \varepsilon }-2\ln \alpha \cdot \frac{\left[ \Gamma (1-\varepsilon
)\right] ^{2}}{\Gamma (1-2\varepsilon )}\left[ 1+\frac{(1+z)^{\varepsilon }-1}{z} 
\right] \right\}.
\end{eqnarray*}
The last contribution (\ref{2.6.8}) is finite and of no interest.


\begin{thebibliography}{99}
\bibitem[\dagger]{foot1} The phrase `free energy' in this problem actually
stands for the grand canonical potential. This terminology is used only to
stress the analogy with $Q$-field theory for free electrons.
\end{thebibliography}
\end{document}